\documentclass[reprint, aps, amsmath, amssymb, superscriptaddress]{revtex4-2}
\usepackage{amsmath}

\setcitestyle{super,open={},close={}}

\usepackage{multirow}
\usepackage{siunitx}

\usepackage{lmodern} 
\usepackage[bookmarks, colorlinks=false]{hyperref}
\usepackage{mathtools}
\hypersetup{colorlinks=true, allcolors=blue}

\begin{document}

\renewcommand{\vec}[1]{\boldsymbol{#1}}

\def\dint#1{\text{d}#1}
\def\grad{\text{grad}}

\newcommand{\comment}[1]{\textcolor{red}{#1}}
\newcommand{\Ed}{E_\mathrm{d}}
\newcommand{\Td}{T_\mathrm{d}}
\newcommand{\Nd}{N_\mathrm{d}}
\newcommand{\kB}{k_\mathrm{B}}
\newcommand{\EA}{E_\mathrm{A}}
\newcommand{\Emelt}{E_\mathrm{melt}}

\hyphenation{milli-dpa}

\title{Microstructure of a heavily irradiated metal exposed to a spectrum of atomic recoils}

\author{Max Boleininger}
	\email{max.boleininger@ukaea.uk}
	\affiliation{UK Atomic Energy Authority, Culham Centre for Fusion Energy, Oxfordshire OX14 3DB, United Kingdom}

\author{Daniel R. Mason}
	\email{daniel.mason@ukaea.uk}
	\affiliation{UK Atomic Energy Authority, Culham Centre for Fusion Energy, Oxfordshire OX14 3DB, United Kingdom}
	
\author{Andrea E. Sand}
	\email{andrea.sand@aalto.fi}
	\affiliation{Department of Applied Physics, Aalto University, 02150 Espoo, Finland}
	
\author{Sergei L. Dudarev}
	\email{sergei.dudarev@ukaea.uk}
	\affiliation{UK Atomic Energy Authority, Culham Centre for Fusion Energy, Oxfordshire OX14 3DB, United Kingdom}

\date{\today}
\begin{abstract}

\noindent At temperatures below the onset of vacancy migration, metals exposed to energetic ions develop dynamically fluctuating steady-state microstructures. Statistical properties of these microstructures in the asymptotic high exposure limit are not universal and vary depending on the energy and mass of the incident ions. We develop a model for the microstructure of an ion-irradiated metal under athermal conditions, where internal stress fluctuations dominate the kinetics of structural evolution. The balance between defect production and recombination depends sensitively not only on the total exposure to irradiation, defined by the dose, but also on the energy of the incident  particles. The model predicts the defect content in the high dose limit as an integral of the spectrum of primary knock-on atom energies, with the finding that low energy ions produce a significantly higher amount of damage than high energy ions.

\end{abstract}
\maketitle

\noindent\textbf{\sffamily INTRODUCTION}\\
\noindent Metals exposed to irradiation by highly energetic particles develop nanoscale structural distortions through the ballistic displacement of atoms out of their crystal lattice sites. In the limit of high exposure, these radiation defects become numerous enough to significantly alter the microstructure, leading to detrimental changes in materials properties, such as reduction in thermal and electrical conductivity, volumetric swelling and dimensional changes, hardening, and embrittlement. The development of a quantitative simulation algorithm for modelling evolving fluctuating microstructures of metals and alloys in a radiation environment has recently acquired prominence in the context of virtual reactor design \cite{Jumel2000,Barrett2018,NF2018, Reali2022}, as failure resulting from the accumulation of radiation damage limits the service life-time of reactor components --- both in the presently operating fission reactors \cite{Odette2009} and in conceptual fusion reactor designs. In particular, it is deemed essential to be able to identify the difference between microstructures produced by exposure to various types of energetic particles, for example neutrons with various energy spectra \cite{Knaster2016} or energetic ions \cite{Was2015} used as cost-effective surrogates for neutron irradiation experiments.


Significant progress has been made recently in the theoretical development and experimental validation of quantitative models predicting the number of defects generated in a perfect crystalline matrix by an individual recoil atom resulting from a collision with an incident high energy particle \cite{Sand2013, nordlund2018improving, yang2021full}. However, these models do not describe the non-linear microstructural changes resulting from the accumulation of defects occurring beyond the low dose limit of exposure of 0.01 displacements per atom (dpa), where the overall density of defects exceeds approximately {$\sim$\,\SI{0.01}{at \%}}. In the low dose limit, the clustering of defects is well described by the power law statistics, discovered in simulations \cite{Sand2013} and confirmed by electron microscope observations \cite{Yi2015}. These power laws are similar to those found in observations of fragmentation of solid projectiles occurring on impact \cite{Oddershede1993}. In a  heavily irradiated metal, the fact that the defects cluster directly in the cascade events appears less significant, since in a material already containing structural distortions, clustering also occurs due to elastic interaction between defects, irrespective of whether they accumulate sequentially or form simultaneously in collision cascades \cite{granberg2016mechanism, Derlet2020, Mason2020}. Eventually at doses above approximately \SI{0.1}{dpa}, clusters of defects start coalescing, and this gives rise to the formation of an extended system-spanning dislocation network that, together with the population of isolated defects continuously produced by irradiation, forms a dynamically fluctuating driven steady state of the material\cite{mason2021parameter, Warwick2021, boleininger2022volume}.

\begin{figure*}[t]
\includegraphics[width=\textwidth]{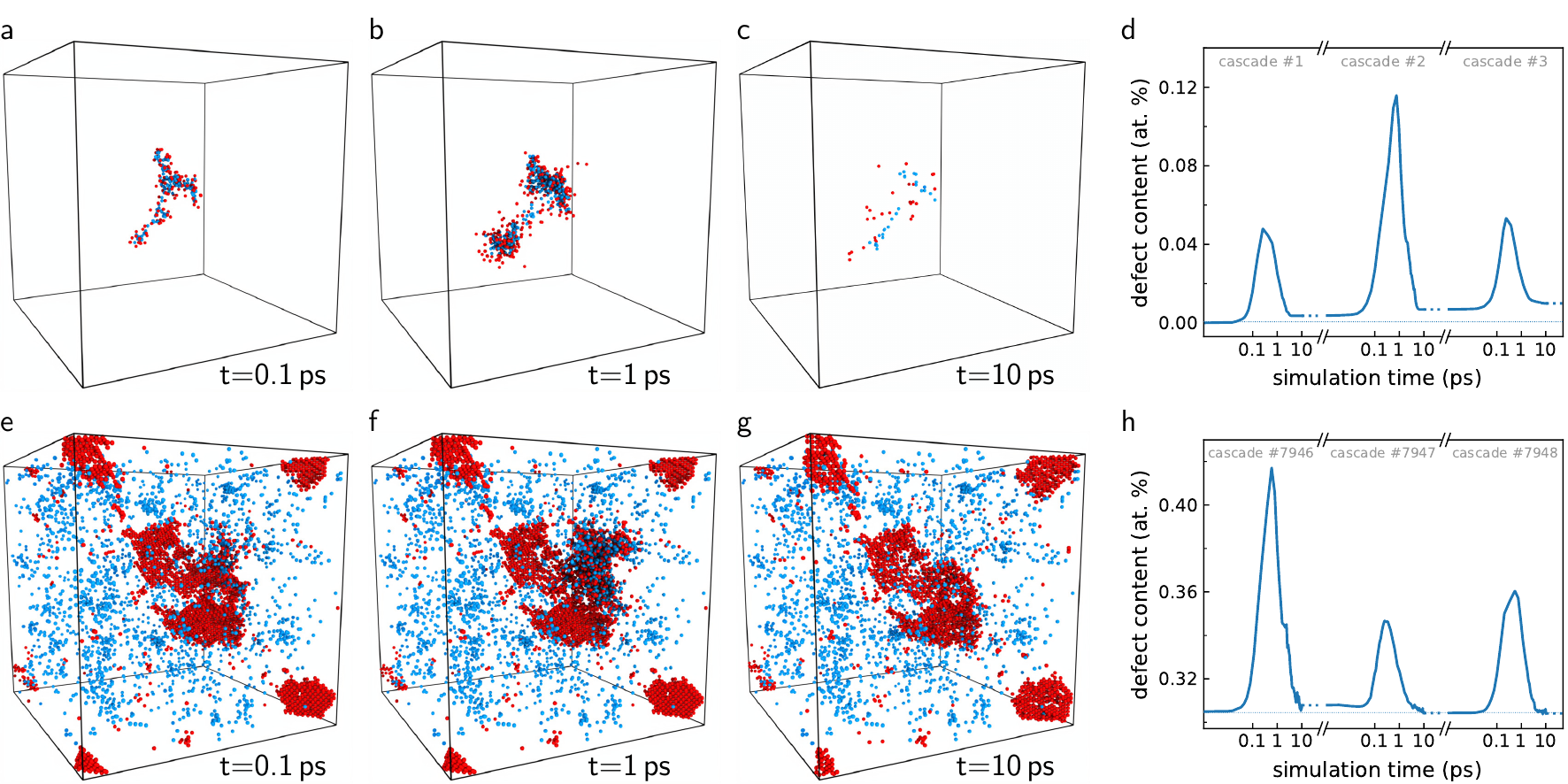}
\caption{\sffamily{
Displacement damage in the driven dynamic steady state of a material exposed to radiation. Damage evolution after initialisation of a cascade with {30\,keV} recoil energy in \textbf{a-c} pristine tungsten and \textbf{e-g} tungsten previously irradiated to {1\,dpa}. Shown are only the crystal defects identified using the Wigner-Seitz defect analysis as interstitials (red) and vacancies (blue). \textbf{d},\textbf{h} Evolution of the defect content produced by cascade initiation.  In pristine crystalline tungsten, the defect content increases linearly with each cascade event, while in the already irradiated tungsten, the defect content fluctuates around a mean value; no new defects are generated as the microstructure is in a steady-state with respect to irradiation.}
\label{fig:snapshots}
}
\end{figure*}

In this study, we develop a quantitative model for predicting the defect content in a heavily irradiated metal, informed by simulations of consecutive cascade impacts extending into the high dose range $\sim$\SI{1}{dpa}. Whilst in general, microstructure is expected to depend on temperature and radiation histories in an intricate and tangled manner, we find that in the athermal limit, where microstructure is driven towards a steady state by defect generation and stress relaxation and not by thermal diffusion, it is possible to formulate a predictive model for the defect content in a heavily irradiated material, analogous to but at the same time fundamentally distinct from models used for predicting defect generation in a perfect crystal lattice. 

We find that radiation drives metals to different dynamic steady states containing  different amounts of defects, depending on the spectrum of recoil energies, and that the defect content can be {\it reversibly} changed by altering the recoil spectrum. We predict that the recoil spectra involving a larger fraction of high energy particles produce microstructures with lower vacancy content, in qualitative and quantitative agreement with experimental assessments of vacancy concentration in tungsten irradiated by high energy ions.

\

\noindent\textbf{\sffamily RESULTS AND DISCUSSION}\\
\noindent\textbf{Properties of the steady state} \\
\noindent The generation of radiation defects is simulated by explicitly propagating atomic trajectories from the moment an incident high energy particle collides with an atom in a crystal lattice, often called the recoil or primary knock-on atom. If sufficient kinetic energy $E_\mathrm{R}$ is transferred to the recoil atom, it is displaced from its lattice site, leading to a cascade that proceeds through three distinct stages of evolution, see Fig.~\ref{fig:snapshots}. In the ballistic phase lasting approximately {$\sim$\,\SI{0.1}{\ps}}, the recoil atom initiates more recoils, which themselves may initiate more recoils, and so on, thereby causing a cascade of self-similar displacements, until no atoms remain with kinetic energy sufficient to create more recoils. This is followed by the heat spike phase of duration {$\sim$\,\SI{1}{\ps}}, caused by the redistribution of the remaining kinetic energy of displaced atoms to the surrounding crystal lattice, which produces a localised region with average atomic energies well over the melting point. In the final cooling phase extending to {$\sim$\,\SI{10}{\ps}}, the ions cool down and recrystallise. The recrystallisation process concludes on a time scale comparable to that of atomic motion, and therefore crystal defects may remain in the volume of the former heat spike --- these are the aforementioned radiation defects.

The state of the art modelling of high dose microstructure at atomic resolution is the sequential simulation of collision cascades\cite{granberg2016mechanism}. Whereas a single collision cascade in a conventional size simulation cell introduces the dose of {$\sim$\,\SI{e-4}{dpa}}\cite{Warwick2021}, higher levels of exposure can be reached by repeating thousands of such cascade simulations successively in the same evolving microstructure. It is computationally feasible to reach high doses of {$\phi$\,$\sim$\,\SI{1}{dpa}} using this algorithm\cite{hirst2022revealing, boleininger2022volume}, however, the total simulated time is unlikely to extend far beyond \SI{1}{\micro s} for system sizes of interest due to the computational limitations of molecular dynamics. The simulated dose rates of {$\mathrm{d}\phi/\mathrm{d}t$\,$\sim$\,\SI{1}{dpa/\micro\second}} are many orders of magnitude above even the comparatively high dose rates of {$\mathrm{d}\phi/\mathrm{d}t$\,$\sim$\,\SI{1}{dpa/hour}} characteristic of ion irradiation experiments. 

Under such extreme dose rates, thermally activated processes that contribute to damage recovery in a simulated microstructure are partly or wholly suppressed because the defects involved are statistically more likely to be reconstituted by a collision cascade before they get a chance to undergo any thermally activated transformation. Reversing the argument, one may also conclude that for a specific thermally-activated process and a given dose rate, there is a temperature below which the process is effectively frozen out  and does not occur, thereby not contributing to microstructural evolution. It is in this \textit{athermal regime} that consecutive cascade simulations offer a valid description of a heavily irradiated material. 

\begin{figure}[t]
\includegraphics[width=\columnwidth]{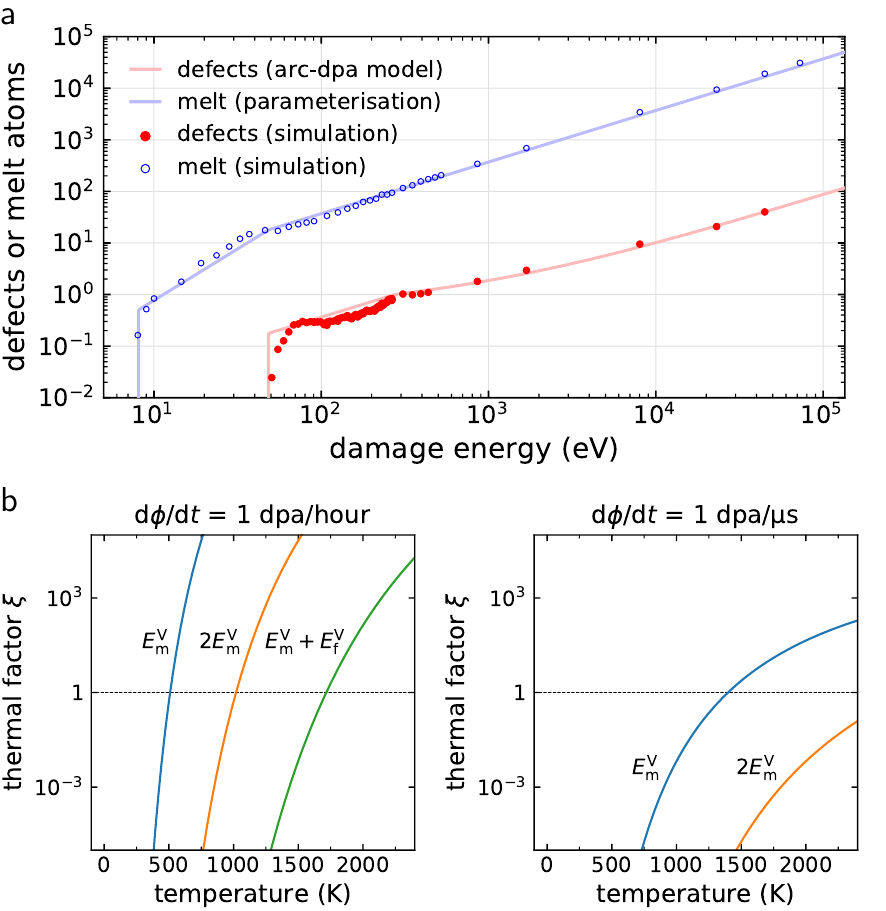}
\caption{\sffamily{
Estimating the phase space of the athermal regime. \textbf{a} Simulation and model for the mean number of defects and molten atoms generated by a cascade with given damage energy in tungsten. The standard error is smaller than the marker size. \textbf{b} Thermal factor \eqref{eq:athermalfactor} for dose rates $\mathrm{d}\phi/\mathrm{d}t$ representative of experiment and of cascade simulations. With decreasing temperature, thermally-activated processes with higher energy barriers become frozen out ($\xi \ll 1$), as any given region in the crystal becomes more likely to be impacted and molten by a cascade before a thermally activated transformation occurs. For pure tungsten, $E_\mathrm{melt} = \SI{2.70}{eV}$ and the transformation barriers are: for vacancy migration $E_m^V = \SI{1.52}{\eV}$, for self-climb\cite{swinburne2016fast} $2E_m^V= \SI{3.04}{\eV}$, and for vacancy-mediated climb\cite{rovelli2017non} $E_m^V+E_f^V = \SI{5.12}{\eV}$.}
\label{fig:timescale}
}
\end{figure}

It is possible to define the athermal regime more precisely for a thermally-activated process characterised by a given activation energy $\EA$, for example vacancy migration. The mean time between the events is given by Arrhenius' law
\begin{equation}\label{eq:timea}
    \tau_\mathrm{A} = \nu_0^{-1} \exp\left(\beta \EA\right),
\end{equation}
where $\nu_0$ is the attempt frequency approximately equal to the Debye frequency divided by $2\pi$. In a system containing $N$ atoms, assume that a collision cascade occurs on average after every time interval $t_\mathrm{c}$.
While the heat spike phase of the cascade does not produce a true equilibrium liquid phase, there is significant mixing possible during this time interval\cite{Nordlund_JNM2018}, and it is a convenient simplification to consider the heat spike as the melting and recrystallisation of $N_\mathrm{melt}$ atoms. In this scenario, the mean time between the successive cascade impacts at any atomic site is given by 
\begin{equation}\label{eq:timec}
    \tau_\mathrm{C} = t_c N/N_\mathrm{melt}.
\end{equation}
According to the formal definition of dose, a single cascade increments it by the ratio of the number of atoms displaced in the ballistic phase $\Nd$ to the system size: $\Delta\phi = \Nd/N$. Eliminating $t_c$ using the expression for dose rate $\dot{\phi} = \Nd/(N t_c)$, we define the \textit{thermal factor} $\xi$ as the ratio of mean times \eqref{eq:timec} to \eqref{eq:timea}:
\begin{equation}
\xi = \frac{\tau_\mathrm{C}}{\tau_\mathrm{A}} = \frac{\nu_0 \Nd}{\dot{\phi} N_\mathrm{melt}}  \exp\left(-\beta \EA \right).
\end{equation}
Next, we use the estimate $\Nd \sim \Td/(2E_d)$\cite{norgett1975proposed}, where {$\Ed\sim$ {10-\SI{100}{\eV}}} is the threshold displacement energy, defined as the average kinetic energy required to displace an atom out of its crystal lattice site averaged over crystallographic directions\cite{nordlund2006molecular}, and the damage energy $\Td < E_\mathrm{R}$ is the remaining kinetic energy after accounting for the energy lost to electronic excitation\cite{lindhard1963range}. Finally, we parameterise the size of the molten cascade region by the phenomenological expression
\begin{equation}\label{eq:meltnumber}
  N_\mathrm{melt}(\Td) =
    \begin{cases}
      \hfil 0,             
                    & \hfil \Td < E_\mathrm{melt}^\mathrm{min} \\
      \hfil \Td^2/(\Ed E_\mathrm{melt}) ,           
                    & \hfil E_\mathrm{melt}^\mathrm{min} \leq \Td < \Ed^\mathrm{min} \\
      \hfil \Td/E_\mathrm{melt},
                    & \hfil \Ed^\mathrm{min} \leq \Td%
    \end{cases},
\end{equation}
where $\Emelt$ is the \textit{effective} energy per atom required to melt the crystal
\footnote{The energy required to melt a crystal with initial temperature $T_0$ is formally obtained by $\Emelt^0 = c_\mathrm{V} (T_\mathrm{melt} - T_0) + L$\cite{de2016subcascade}, 
where $c_\mathrm{V} \approx 3 k_\mathrm{B}$ is molar heat capacity, $T_\mathrm{melt}$ is melting temperature, and $L$ is latent heat of fusion per atom. In practice this value overestimates the melt size as it does not account for kinetic energy being rapidly conducted out of the cascade region into the surrounding crystal.}%
, as obtained from fits to single cascade simulations similarly to defect production models\cite{nordlund2018improving}, see Fig.~\ref{fig:timescale}a, and $\Ed^\mathrm{min}$ is the threshold displacement energy minimum with respect to crystallographic directions. For the damage energy region of interest $\Td > \Ed$, the thermal factor can be simplified to
\begin{equation}\label{eq:athermalfactor}
\xi \approx \frac{\nu_0 \Emelt}{2\dot{\phi} \Ed}  \exp\left(-\beta \EA \right).
\end{equation}

The thermal factor is the mean number of times a defect takes part in a thermal migration event before being enveloped by the next cascade. In the athermal regime {$\xi$\,$\ll$\,1}, a defect is much more likely to be reconstituted by a cascade than undergo a thermal hop, while in the thermal regime {$\xi$\,$\gg$\,1}, a defect is able to migrate many times, transferring mass through the system or coalescing with other defects, before being embroiled in another cascade. Examples of thermal factors for various processes in pure tungsten are shown in Fig.~\ref{fig:timescale}. For vacancies migrating in pure tungsten, equation (\ref{eq:athermalfactor}) shows that {$\xi$\,$\ll$\,1} at room temperature and dose rates representative of ion irradiation experiments, suggesting that consecutive cascade simulation indeed offers a valid description of microstructural evolution under these conditions. 

\

\begin{figure}[htb!]
\includegraphics[width=\columnwidth]{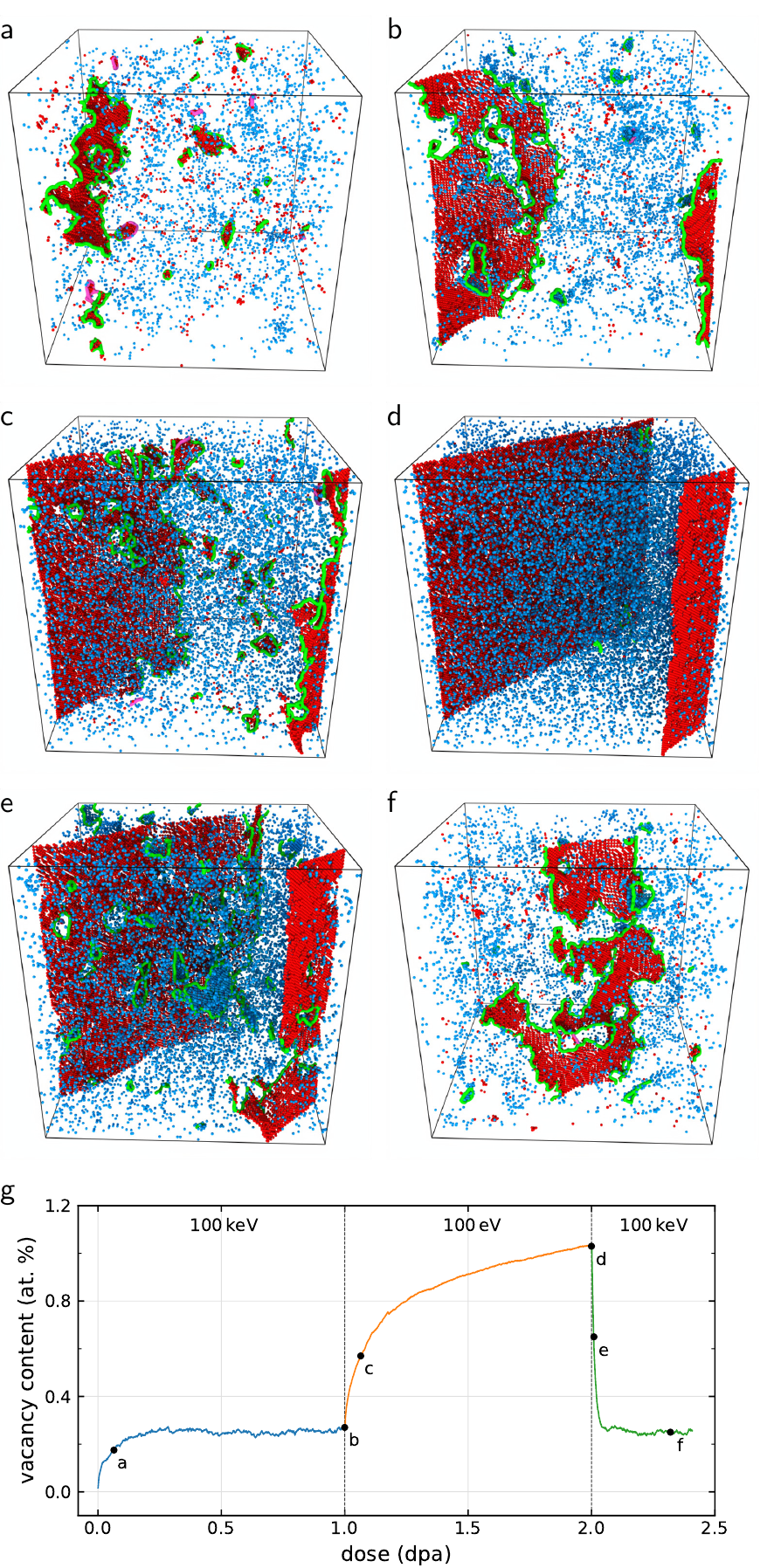}
\caption{\sffamily{
Reversibility of driven high-dose microstructure. \textbf{a-b} Pristine tungsten is damaged by {100\,keV} cascades until saturation. \textbf{c-d} Switching to {100\,eV} cascades drives the system towards a steady state with higher vacancy content. \textbf{e-f} Switching back to {100\,keV} cascades returns the system to the previous steady state with lower vacancy content through formation and coalescence of vacancy loops.
Green (pink) lines show the position of $1/2\langle 111 \rangle$ ($\langle 100 \rangle$) dislocation lines detected using the DXA method \cite{stukowski2009visualization}.
}
\label{fig:flip}
}
\end{figure}

\

\noindent\textbf{Defect content in a driven steady state}\\
\noindent In the athermal limit, where all the thermally-activated processes are frozen out, irradiated microstructure develops through the balance between defect generation by cascades and defect recombination during the recrystallisation of cascade heat spikes. Considering the fact that irradiation by highly energetic particles generates a broad spectrum of recoil energies, one poses a question if the damage produced by the cascades with a given recoil energy can be reversed by cascades with a different damage energy, or whether the crystal structure retains memory of its irradiation history.

Figure~\ref{fig:flip} illustrates microstructural evolution of the initially pristine tungsten that is successively damaged to saturation first by $\SI{100}{\keV}$ energy cascades, followed by $\SI{100}{\eV}$ energy cascades, and finally again by $\SI{100}{\keV}$ energy cascades. The vacancy content, shown in figure~\ref{fig:flip}g, is found to saturate to different levels depending on the recoil energy: switching from \SI{100}{keV} to \SI{100}{eV} cascades drives the vacancy content from {0.32\,at.\,\%} to {$\sim$1\,at.\,\%}, after which switching back to \SI{100}{keV} cascades drives the vacancy content back to {0.32\,at.\,\%}. The final vacancy content is such as if the \SI{100}{eV} cascade exposure had not occurred at all. 

We observe that the vacancy content saturation to a steady state is entirely determined by the recoil energy, irrespective of the irradiation history. While the migration barrier for some interstitial-type defects is sufficiently small to allow for their athermal migration \cite{Dudarev2008}, driven by the fluctuating stress fields developing in the irradiated microstructure \cite{Derlet2020}, vacancies and vacancy clusters are effectively immobile in the athermal limit of microstructural evolution, remaining stationary between the successive cascade impacts. In the high dose limit, every part of the initially crystalline material has molten and recrystallised at least once, with the average cascade impacts generating a fixed number of vacancies and a fixed mean number of recrystallised atoms, depending on the damage energy. 

Based on the above reasoning, we propose the following expression for the average vacancy concentration $c_\mathrm{V}^\mathrm{sat}$ in the athermal steady state driven by collision cascades with damage energy $\Td$:
\begin{equation}\label{eq:csatformal}
    c_\mathrm{V}^\mathrm{sat}(\Td) = \frac{N_\mathrm{V}^\mathrm{sat}(\Td)}{N_\mathrm{melt}^\mathrm{sat}(\Td)},
\end{equation}
where $N_\mathrm{V}^\mathrm{sat}(\Td)$ is the number of vacancies remaining after recrystallisation of $N_\mathrm{melt}^\mathrm{sat}(\Td)$ atoms in the steady-state microstructure. These quantities are essentially unknown, and their determination would require  extensive analysis of collision cascades in the already saturated microstructures. However, under the assertion that the microstructure of the athermal steady state is qualitatively well described as a crystal with highly elevated vacancy content, see figure~\ref{fig:flip}d, the models for cascades in a perfect crystal, surprisingly, might also be able to describe the saturated steady state. Using the above expression for the size of the cascade melt in an initially pristine crystal \eqref{eq:meltnumber}, we also make use of the athermal recombination corrected displacement model (arc-dpa) introduced by Nordlund \textit{et al.}\cite{nordlund2018improving}, in the form modified by Yang and Olsson\cite{yang2021full}, to obtain an estimate for the number of vacancies generated by a single cascade:
\begin{equation}\label{eq:arcnumber}
  \Nd^\mathrm{arc}(\Td) =
    \begin{cases}
      \hfil 0,             
                    & \hfil \Td < \Ed^\mathrm{min} \\
      \hfil \frac{0.8 \Td}{2 \Ed},           
                    & \hfil \Ed^\mathrm{min} \leq \Td < \frac{2\Ed}{0.8} \\
      \hfil \frac{0.8 \Td}{2 \Ed} \xi(\Td),
                    & \hfil \frac{2\Ed}{0.8} \leq \Td%
    \end{cases},
\end{equation}
where function $\xi(\Td) = (1 - c)\left(0.8\Td/(2\Ed)\right)^b + c$ is a phenomenological expression accounting for the sub-linear defect production at intermediate damage energies, with parameters $b$ and $c$ fitted such that $\Nd^\mathrm{arc}$ accurately matches the number of defects obtained from molecular dynamics simulations for a given material \cite{konobeyev2017evaluation, nordlund2018improving, yang2021full}. 

The central outcome of this analysis is an approximation for the formally unknown number of vacancies and recrystallised atoms in the saturated regime by their counterparts in a perfect crystal: $N_\mathrm{V}^\mathrm{sat} \approx \Nd^\mathrm{arc}$ and $N_\mathrm{melt}^\mathrm{sat} \approx N_\mathrm{melt}$. Hence, we arrive at a simple expression for predicting the saturated vacancy concentration $c^\mathrm{sat}_\mathrm{V}$ in a high-dose microstructure
\begin{equation}\label{eq:csat}
    c^\mathrm{sat}_\mathrm{V}(\Td) \approx \frac{\Nd^\mathrm{arc}(\Td)}{N_\mathrm{melt}(\Td)},
\end{equation}
where $N_\mathrm{melt}$ and $\Nd^\mathrm{arc}$ are given by Eqs.~\eqref{eq:meltnumber} and \eqref{eq:arcnumber}, respectively. Expression \eqref{eq:csat} is equivalent to stating that in a heavily irradiated material in the athermal limit, the history of microstructural evolution is completely irrelevant for the vacancy type defects. We do not discuss here the self-interstitials, because while formally the vacancy and interstitial counts must be equal, interstitials exhibit a strong drive towards coalescence\cite{mason2019relaxation}, and the eventual formation of extended dislocation networks\cite{Derlet2020} or even complete crystal planes\cite{Mason2020, boleininger2022volume}, and as such may not be readily recognisable as point defects.

\begin{figure}[t]
\includegraphics[width=\columnwidth]{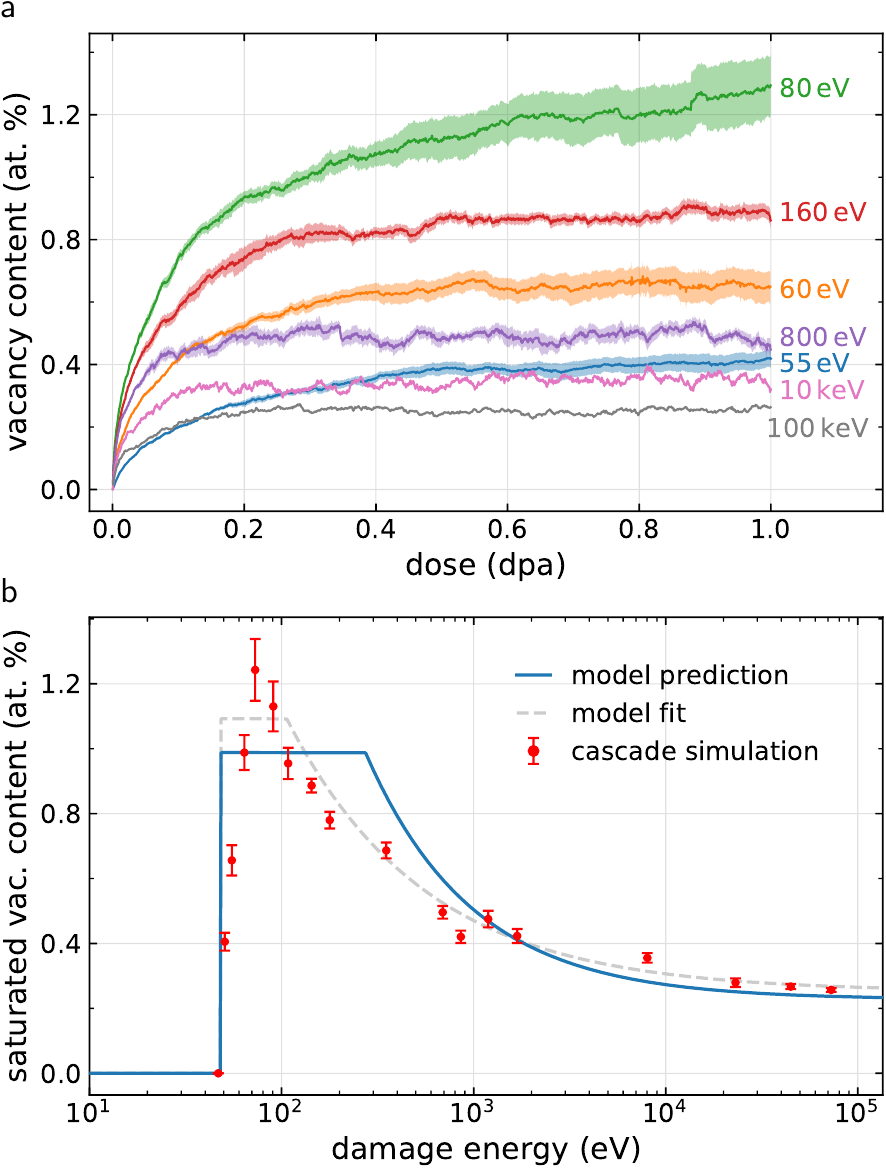}
\caption{\sffamily{}
Vacancy content as a function of recoil energy. \textbf{a} The vacancy concentration in tungsten irradiated up to a dose of {1\,dpa}, as obtained from cascade simulations, is found to decrease with increasing recoil energy. Shaded areas indicate the standard error obtained from multiple simulations. \textbf{b} Comparison of the vacancy concentration from simulation in the {$\sim${1\,dpa}} high dose limit (points) with the analytic prediction (line) given by the number density of vacancies generated in a cascade in an initially pristine crystal. Fitting the model by improves quantitative agreements (dashed), but removes the predictive aspect of the model. Mean values are averaged over {0.9-1.0\,dpa}, with error bars showing the standard deviation, or in the case of multiple simulations, the mean standard error, over the same interval.
\label{fig:distributepka}
}
\end{figure}

We have validated the vacancy concentration model \eqref{eq:csat} by running cascade simulations in tungsten up to the dose of $\SI{1}{dpa}$ over a broad range of recoil energies. We find that vacancy concentration saturates well below the dose of $\SI{1}{dpa}$, see Fig.~\ref{fig:distributepka}a, in agreement with independent simulations  \cite{granberg2016mechanism, Derlet2020, mason2021parameter} and experimental observations \cite{tian2021heavy, zhang2021comparing}. This can be rationalised by recognising that the number of molten atoms per cascade is much higher than the number of ballistically displaced atoms, as suggested by their respective characteristic energy scales $\Td/\Emelt \gg \Td/\Ed$. In Fig.~\ref{fig:distributepka}b we compare the saturated vacancy concentration with the analytical prediction \eqref{eq:csat}, and find that the prediction is in qualitative and quantitative agreement with the simulation results. Most notably, we find that impacts with smaller recoil energies that are characteristic of irradiation by light particles, produce a significantly higher vacancy content than impacts with high recoil energies, characteristic of irradiation by heavy particles. 

Predictions of the athermal saturated vacancy content can be made for a variety of materials, as parameters for $\Nd^\mathrm{arc}$ are available in literature for a broad selection of metals and alloys \cite{konobeyev2017evaluation, nordlund2018improving, yang2021full}, and as $\Emelt$ can either be obtained from cascade simulations or estimated. We have  tested the transferability of the model using face-centered cubic (fcc) copper and hexagonal close-packed (hcp) zirconium for which we performed cascade simulations using \SI{100}{eV}, \SI{1}{keV}, and \SI{10}{keV} recoil energies up to \SI{1}{dpa}, see table~\ref{tab:transferability}. The predictions are in qualitative agreement with data from high dose cascade simulations. In principle, a few data points from high dose cascade simulations could be used to adjust the model parameters for better quantitative agreement, as shown in Fig.~\ref{fig:distributepka}b. However, we emphasise that high dose simulations require several orders of magnitude more computational resources than the single-cascade simulations used for parameterising the predictive model; generating the single-cascade data shown in Fig.~\ref{fig:timescale} required only $\sim$5,000 core-hours, compared to $\sim$1,000,000 core-hours required for accumulating the high dose data shown in Fig.~\ref{fig:distributepka}.

\begin{table}[t]
{\sffamily
  \caption{
Saturated vacancy content in {at.\,\%} for three types of crystal structures damaged by cascades with given recoil energy. Comparison between prediction \eqref{eq:csat} and simulation at {1\,dpa} shows that the model qualitatively reproduces the drop in vacancy content as the recoil energy increases.\label{tab:transferability}}
\begin{ruledtabular}  
\begin{tabular}{clccc}
material & method & {100\,eV} & {1\,keV} & {10\,keV} \\
   \hline
    bcc-W\cite{mason2017empirical}  
        &    simulation  & 1.11\,$\pm$\,0.07 & 0.56\,$\pm$\,0.09 & 0.40\,$\pm$\,0.03 \\
        &    prediction  & 1.02 & 0.62 & 0.34\\[.3em]
    fcc-Cu\cite{ackland1990many} 
        &    simulation  & 1.82\,$\pm$\,0.02 & 0.70\,$\pm$\,0.01 & 0.22\,$\pm$\,0.02 \\
        &    prediction  & 1.54 & 0.73 & 0.36 \\[.3em]
    hcp-Zr\cite{Mendelev2007} 
        &    simulation  & 0.97\,$\pm$\,0.03 & 0.49\,$\pm$\,0.01 & 0.27\,$\pm$\,0.02 \\
        &    prediction  & 0.99 & 0.58 & 0.42
  	\end{tabular}
\end{ruledtabular}
}
\end{table}

In the general case where there is a distribution of damage energies $p(\Td)$, the probability for an atom in the crystal to become part of a heat spike of an incident cascade is equal to $p_\mathrm{hit}(\Td) = p(\Td) N^\mathrm{sat}_\mathrm{melt}(\Td)/N$, where $N$ is the total number of atoms. The mean saturated vacancy concentration, averaged across a volume region of interest, is therefore obtained as the expectation value of the vacancy concentration $c_\mathrm{V}^\mathrm{sat}(\Td)$ over the distribution $p_\mathrm{hit}(\Td)$, resulting in the expression
\begin{equation}\label{eq:csatmean}
    \left\langle c_\mathrm{V}^\mathrm{sat} \right\rangle = \frac{
    \int_{E_\mathrm{d}^\mathrm{min}}^\infty N_\mathrm{V}^\mathrm{sat}(\Td) p(\Td)\, \dint{\Td}
    }{
    \int_{E_\mathrm{d}^\mathrm{min}}^\infty N^\mathrm{sat}_\mathrm{melt}(\Td) p(\Td)\, \dint{\Td}
    }
\end{equation}
after some simplification using Eq.~\eqref{eq:csatformal}.

\begin{figure*}[t]
\includegraphics[width=\textwidth]{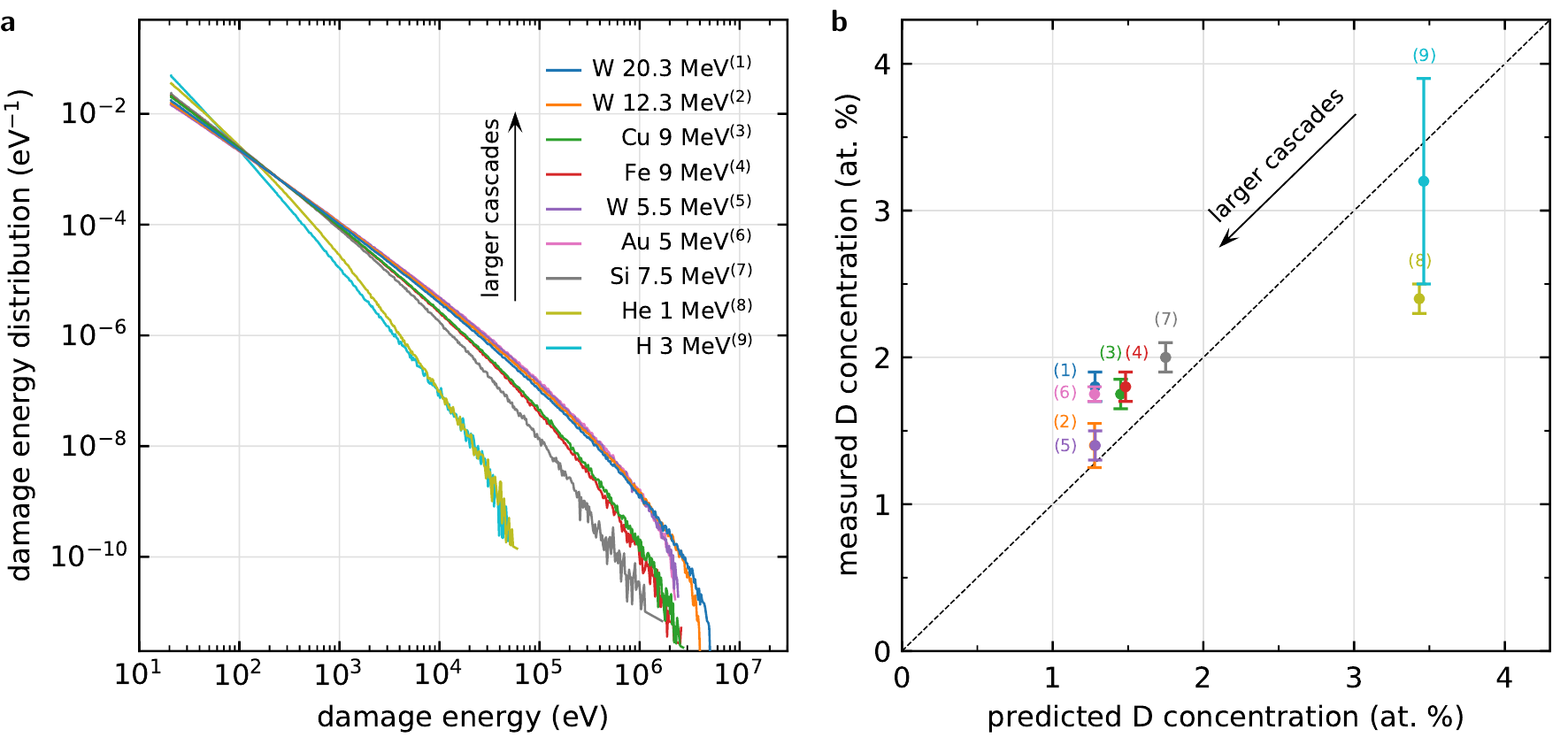}
\caption{\sffamily{
Saturated deuterium content in the high dose limit. \textbf{a} Distribution of cascade damage energies for various irradiation ions. Ions with higher momentum are more likely to generate higher energy cascades. \textbf{b} Comparison of analytically predicted \eqref{eq:csatmean} and experimentally measured deuterium concentrations in the saturated limit. Irradiation with lower momentum ions leads to a higher deuterium content, consistent with the higher vacancy content produced by smaller cascades. Colors and labels are consistent with damage energy spectra from Fig.~\ref{fig:expcomparision}\textbf{a}. Measured concentrations extracted from (6)\cite{zhang2021comparing}, (1,3,4,7,8)\cite{wielunska2020deuterium}, (5)\cite{tyburska2009deuterium}, (2)\cite{t2012saturation}, (9)\cite{moller2020deuterium}.}
\label{fig:expcomparision}
}
\end{figure*}

\

\noindent\textbf{\sffamily Experimental validation}\\
\noindent As performing a direct quantitative measurement of vacancy content in an irradiated material is highly non-trivial \cite{Meslin2010}, we opted for an indirect comparison to depth-resolved deuterium concentration measurements. We consider deuterium retention in irradiated tungsten as, due to the expected use of tungsten for plasma-facing components in fusion reactor designs, the data derived from such experiments are available in the literature over a broad variety of irradiation conditions. In the deuterium retention experiments considered here, the material is first damaged by ion irradiation and then exposed to a deuterium plasma at elevated temperatures $T \sim \SI{400}{\kelvin}$ over the course of several days. As the deuterium particle energies are much smaller than the threshold displacement energy, the deuterium is only implanted into a shallow surface layer from where it subsequently diffuses into the irradiation-damaged layer of the material whilst binding to trapping sites. First-principles simulations found that the deuterium binds preferentially to void surfaces, with a single tungsten vacancy trapping up to 5 deuterium atoms with trapping energies above \SI{1.1}{\eV} \cite{heinola2010hydrogen}.

We compiled nuclear reaction analysis measurements of peak deuterium concentrations in tungsten irradiated to doses $\phi \gtrsim \SI{0.3}{dpa}$ under athermal condition $T<\SI{500}{\kelvin}$ for a broad range of irradiation energies from a number of studies \cite{wielunska2020deuterium, tyburska2009deuterium, zhang2021comparing, t2012saturation, moller2020deuterium}. For each particle energy represented in the experimental data, we simulated the recoil spectra, see Fig.~\ref{fig:expcomparision}, and used them to compute the expected saturated vacancy concentration using Eq.~\eqref{eq:csatmean}.  In Fig.~\ref{fig:expcomparision}b we compare the predicted deuterium concentrations computed as $\left\langle c_\mathrm{D}^\mathrm{sat}\right\rangle = 5 \left\langle c_\mathrm{V}^\mathrm{sat}\right\rangle$ \cite{mason2021parameter} with the experimentally measured deuterium concentration in the saturation regime. The predictions closely follow the trends found in experiment. We find that ions that have a higher probability of generating smaller cascades also generate microstructures characterised by higher deuterium retention, which affirms the prediction that low energy cascades correspond to higher  large-dose-microstructure vacancy concentrations, see Fig.~\ref{fig:distributepka}. We also computed the deuterium concentrations using the fitted model shown in Fig.~\ref{fig:distributepka}b, and found that the results differ only by a mean relative error of \SI{7}{\percent}. The fitted model does offer better accuracy, see the Supplemental Material, but also requires data from computationally intensive high dose simulations, which on balance appear unwarranted since there are other comparatively less controlled sources of uncertainty present here. 

Although the experiments \cite{wielunska2020deuterium, tyburska2009deuterium, zhang2021comparing, t2012saturation, moller2020deuterium} were conducted at temperatures below which vacancies become mobile, referred to as stage III recovery \cite{keys1968high}, some lower-barrier interstitial migration processes may still contribute towards the interstitial-vacancy recombination. We do not expect this effect to play a significant role in tungsten, where interstitial migration barriers are so low that they are driven by the fluctuating stress fields present in the microstructure under irradiation. An additional source of systematic error is our assumption that the entire vacancy-type content produced in a cascade is represented solely in the form of mono-vacancies, ignoring the possibility of vacancy clustering. From the comparison with our high dose cascade overlap simulations, we find that even with the uncertainty introduced by using an empirical interatomic potential, we can still estimate the deuterium content to around 10\% of the experimentally measured value.

Deuterium retention experiments for neutron-irradiated tungsten under athermal conditions are comparatively scarce. Shimada \textit{et al.}\cite{shimada2014irradiation} report a value of \SI{0.8}{\percent} retention for tungsten irradiated to \SI{0.3}{dpa} at \SI{70}{\celsius} in the High Flux Isotope Reactor (HFIR) reactor, with deuterium subsequently implanted at \SI{200}{\celsius}. Using a recoil spectrum representative of HFIR\cite{huang2018simulating}, we predict a deuterium concentration of \SI{1.5}{\percent}. It is encouraging that the prediction overestimates the retention, given that the experiment was conducted at a temperature where some extent of defect recombination through vacancy migration is expected.

\

\noindent\textbf{\sffamily CONCLUSION AND OUTLOOK}\\
\noindent The exposure of metals to high dose irradiation at relatively low temperatures eventually leads to a saturation of defect content. The actual magnitude of this content depends on the primary recoil energy, with low energy cascades generating up to 5 times higher vacancy concentrations than high energy cascades. Our central result is that the number density of vacancies generated within the characteristic volume of the cascade can be used as a predictor for the saturated vacancy concentration, valid over many orders of magnitude in recoil energy spectra. The model is transferable across different metals, provided that the irradiation conditions are athermal, meaning that the dose rate is high relative to the rate of thermal relaxation of defect configurations. The model prediction is validated by comparison to experimental data on the saturated deuterium concentration in highly irradiated tungsten over a broad range of ion energies and masses. 

The saturated vacancy concentration acts as a single number quantifying radiation damage in simplest terms, and can be used to predict the outcome of irradiation with different particle energies and masses \textit{a priori}. This is analogous to how total exposure to radiation is measured in units of displacements per atoms, transferable across different materials and computed from the entire recoil spectrum. While the prediction of the saturated damage state is formally valid in the particular regime of irradiation conditions where thermal defect migration is negligible, this prediction further serves as a rigorously defined baseline for characterising deviations from the steady-state, for instance when thermal diffusion is active.

\

\noindent\textbf{\sffamily METHODS}\\
{\footnotesize
\textbf{\sffamily Molecular dynamics cascade simulations.} MD simulations were run using \textsc{lammps} \cite{plimpton1995fast} with interatomic interactions described by empirical potentials of the embedded atom model type: W (Ref: \cite{mason2017empirical}), Cu (Ref: \cite{ackland1990many}), and Zr (Ref: \cite{Mendelev2007}). Simulation cells were initialised with perfect crystal structure, with periodic boundary conditions applied to all three directions and box dimensions chosen consistent to the {0\,K} equilibrium lattice constant. Starting with an initial dose of {0\,dpa}, irradiation damage is introduced following an iterative procedure: First, a small number of atoms $N_\mathrm{C}$ are selected at random and their kinetic energy set to the recoil energy, with randomly-oriented velocities. If $N_\mathrm{C}>1$, we avoided selecting random atoms too close to one another in order to avoid spurious effects caused by overlapping cascades. Next, the simulation is propagated in the NVE ensemble including some damping terms, see below, until the crystal cools below {100\,K} and at least {5\,ps} are simulated, after which velocities are zeroed out and atomic coordinates relaxed to a local energy minimum using the method of conjugate gradients. This procedure is repeated until a target dose is reached, with each repetition incrementing the dose by $0.8\Td N_\mathrm{C}/(2\Ed N)$, where $N$ is the number of atoms in the simulation cell. We chose $N_\mathrm{C}$ such that the dose is incremented by {$\sim${0.0002\,dpa}} per iteration as a compromise between computational efficiency and avoiding excessive heating of the crystal. System sizes varied between 65,000 and 2,000,000 atoms, with smaller cells and multiple repetitions used for lower energy recoils. The vacancy content was determined using the method described in Ref.\cite{mason2021parameter}. Microstructural images were rendered with \textsc{ovito}\cite{stukowski2009visualization}.

Energy loss to electronic excitation was included by adding a frictional force of magnitude $mv/\tau$ to all atoms with kinetic energy higher than {10\,eV}, where $m$ is the atomic mass and $\tau$ is the time damping constant extracted from the
low velocity self-ion stopping limit of the \textsc{srim} code\cite{ziegler2004srim}. Electronic stopping time constants are {0.90\,ps}, {1.04\,ps}, and {2.37\,ps} for W, Cu, and Zr, respectively. For atoms with temperature below their melting temperature, another damping term was added to model energy loss due to electron-phonon coupling\cite{mason2015incorporating} with time constants of {16.0\,ps}, {26.4\,ps}, and {12.5\,ps} for W, Cu, and Zr, respectively. We refer to the Supplemental Material for more information on how these parameters were obtained.

Parameters for the arc-dpa and melt models for the interatomic potentials used here were obtained from single cascade simulations and are listed in the Supplemental Material. The size of the melt was determined using \textsc{ovito}\cite{stukowski2009visualization} by filtering out atoms of the stable crystal structure type as determined by the common neighbour analysis method, after which atoms with less than $3/4$ of the equilibrium coordination number are filtered out to remove the crystalline region surrounding the melt. The remaining atoms have a liquid-like radial distribution function. The melt size is then obtained as the maximum size of the melt over the course of a cascade.

\

\noindent\textbf{\sffamily Recoil spectra simulations.} Recoil energies were obtained using the \textsc{srim} software\cite{ziegler2004srim}, which is based on the binary collision approximation method. Recoil energies were converted into damage energies using the Lindhard stopping formula\cite{lindhard1963range} as described in Ref:\cite{norgett1975proposed}. The damage energies were binned, and bin counts divided by the bin width and by the total number of recoils, to obtain a discrete representation of the normalised probability distribution $p(\Td)$ for a recoil to have the damage energy $\Td$, shown in Fig.~\ref{fig:expcomparision}a. The recoil spectrum for the HFIR reactor is shown in the Supplemental Material.
}

\

\noindent\textbf{\sffamily  DATA AVAILABILITY}\\
{\footnotesize
Files containing cascade statistics and the cascade overlap structure files that where produced using molecular dynamics simulations are freely available for download at [TBD].
}

\bibliography{main.bbl}

\begin{thebibliography}{49}%
\makeatletter
\providecommand \@ifxundefined [1]{%
 \@ifx{#1\undefined}
}%
\providecommand \@ifnum [1]{%
 \ifnum #1\expandafter \@firstoftwo
 \else \expandafter \@secondoftwo
 \fi
}%
\providecommand \@ifx [1]{%
 \ifx #1\expandafter \@firstoftwo
 \else \expandafter \@secondoftwo
 \fi
}%
\providecommand \natexlab [1]{#1}%
\providecommand \enquote  [1]{``#1''}%
\providecommand \bibnamefont  [1]{#1}%
\providecommand \bibfnamefont [1]{#1}%
\providecommand \citenamefont [1]{#1}%
\providecommand \href@noop [0]{\@secondoftwo}%
\providecommand \href [0]{\begingroup \@sanitize@url \@href}%
\providecommand \@href[1]{\@@startlink{#1}\@@href}%
\providecommand \@@href[1]{\endgroup#1\@@endlink}%
\providecommand \@sanitize@url [0]{\catcode `\\12\catcode `\$12\catcode
  `\&12\catcode `\#12\catcode `\^12\catcode `\_12\catcode `\%12\relax}%
\providecommand \@@startlink[1]{}%
\providecommand \@@endlink[0]{}%
\providecommand \url  [0]{\begingroup\@sanitize@url \@url }%
\providecommand \@url [1]{\endgroup\@href {#1}{\urlprefix }}%
\providecommand \urlprefix  [0]{URL }%
\providecommand \Eprint [0]{\href }%
\providecommand \doibase [0]{https://doi.org/}%
\providecommand \selectlanguage [0]{\@gobble}%
\providecommand \bibinfo  [0]{\@secondoftwo}%
\providecommand \bibfield  [0]{\@secondoftwo}%
\providecommand \translation [1]{[#1]}%
\providecommand \BibitemOpen [0]{}%
\providecommand \bibitemStop [0]{}%
\providecommand \bibitemNoStop [0]{.\EOS\space}%
\providecommand \EOS [0]{\spacefactor3000\relax}%
\providecommand \BibitemShut  [1]{\csname bibitem#1\endcsname}%
\let\auto@bib@innerbib\@empty
\bibitem [{\citenamefont {Jumel}\ \emph {et~al.}(2000)\citenamefont {Jumel},
  \citenamefont {Domain}, \citenamefont {Ruste}, \citenamefont {{Van Duysen}},
  \citenamefont {Becquart}, \citenamefont {Legris}, \citenamefont {Pareige},
  \citenamefont {Barbu},\ and\ \citenamefont {Pontikis}}]{Jumel2000}%
  \BibitemOpen
  \bibfield  {author} {\bibinfo {author} {\bibfnamefont {S.}~\bibnamefont
  {Jumel}}, \bibinfo {author} {\bibfnamefont {C.}~\bibnamefont {Domain}},
  \bibinfo {author} {\bibfnamefont {J.}~\bibnamefont {Ruste}}, \bibinfo
  {author} {\bibfnamefont {J.~C.}\ \bibnamefont {{Van Duysen}}}, \bibinfo
  {author} {\bibfnamefont {C.}~\bibnamefont {Becquart}}, \bibinfo {author}
  {\bibfnamefont {A.}~\bibnamefont {Legris}}, \bibinfo {author} {\bibfnamefont
  {P.}~\bibnamefont {Pareige}}, \bibinfo {author} {\bibfnamefont
  {A.}~\bibnamefont {Barbu}},\ and\ \bibinfo {author} {\bibfnamefont
  {V.}~\bibnamefont {Pontikis}},\ }\bibfield  {title} {\bibinfo {title}
  {Simulation of the irradiation effects in reactor materials: The {REVE}
  project},\ }\href {https://doi.org/10.1051/jp4:2000633} {\bibfield  {journal}
  {\bibinfo  {journal} {J. Phys. IV France}\ }\textbf {\bibinfo {volume}
  {10}},\ \bibinfo {pages} {Pr6} (\bibinfo {year} {2000})}\BibitemShut
  {NoStop}%
\bibitem [{\citenamefont {Barrett}\ \emph {et~al.}(2018)\citenamefont
  {Barrett}, \citenamefont {Evans}, \citenamefont {Fursdon}, \citenamefont
  {Domptail}, \citenamefont {McIntosh}, \citenamefont {Iglesias},\ and\
  \citenamefont {Surrey}}]{Barrett2018}%
  \BibitemOpen
  \bibfield  {author} {\bibinfo {author} {\bibfnamefont {T.~R.}\ \bibnamefont
  {Barrett}}, \bibinfo {author} {\bibfnamefont {L.~M.}\ \bibnamefont {Evans}},
  \bibinfo {author} {\bibfnamefont {M.}~\bibnamefont {Fursdon}}, \bibinfo
  {author} {\bibfnamefont {F.}~\bibnamefont {Domptail}}, \bibinfo {author}
  {\bibfnamefont {S.~C.}\ \bibnamefont {McIntosh}}, \bibinfo {author}
  {\bibfnamefont {D.}~\bibnamefont {Iglesias}},\ and\ \bibinfo {author}
  {\bibfnamefont {E.}~\bibnamefont {Surrey}},\ }\bibfield  {title} {\bibinfo
  {title} {Virtual engineering of a fusion reactor: application to divertor
  design, manufacture, and testing},\ }\href
  {https://doi.org/10.1109/TPS.2018.2856888} {\bibfield  {journal} {\bibinfo
  {journal} {IEEE Transactions on Plasma Science}\ }\textbf {\bibinfo {volume}
  {47}},\ \bibinfo {pages} {889} (\bibinfo {year} {2018})}\BibitemShut
  {NoStop}%
\bibitem [{\citenamefont {Dudarev}\ \emph {et~al.}(2018)\citenamefont
  {Dudarev}, \citenamefont {Mason}, \citenamefont {Tarleton}, \citenamefont
  {Ma},\ and\ \citenamefont {Sand}}]{NF2018}%
  \BibitemOpen
  \bibfield  {author} {\bibinfo {author} {\bibfnamefont {S.~L.}\ \bibnamefont
  {Dudarev}}, \bibinfo {author} {\bibfnamefont {D.~R.}\ \bibnamefont {Mason}},
  \bibinfo {author} {\bibfnamefont {E.}~\bibnamefont {Tarleton}}, \bibinfo
  {author} {\bibfnamefont {P.-W.}\ \bibnamefont {Ma}},\ and\ \bibinfo {author}
  {\bibfnamefont {A.~E.}\ \bibnamefont {Sand}},\ }\bibfield  {title} {\bibinfo
  {title} {A multi-scale model for stresses, strains and swelling of reactor
  components under irradiation},\ }\href
  {https://doi.org/10.1088/1741-4326/aadb48} {\bibfield  {journal} {\bibinfo
  {journal} {Nuclear Fusion}\ }\textbf {\bibinfo {volume} {58}},\ \bibinfo
  {pages} {126002} (\bibinfo {year} {2018})}\BibitemShut {NoStop}%
\bibitem [{\citenamefont {Reali}\ \emph {et~al.}(2022)\citenamefont {Reali},
  \citenamefont {Boleininger}, \citenamefont {Gilbert},\ and\ \citenamefont
  {Dudarev}}]{Reali2022}%
  \BibitemOpen
  \bibfield  {author} {\bibinfo {author} {\bibfnamefont {L.}~\bibnamefont
  {Reali}}, \bibinfo {author} {\bibfnamefont {M.}~\bibnamefont {Boleininger}},
  \bibinfo {author} {\bibfnamefont {M.~R.}\ \bibnamefont {Gilbert}},\ and\
  \bibinfo {author} {\bibfnamefont {S.~L.}\ \bibnamefont {Dudarev}},\
  }\bibfield  {title} {\bibinfo {title} {Macroscopic elastic stress and strain
  produced by irradiation},\ }\href
  {https://doi.org/https://doi.org/10.1088/1741-4326/ac35d4} {\bibfield
  {journal} {\bibinfo  {journal} {Nuclear Fusion}\ }\textbf {\bibinfo {volume}
  {62}},\ \bibinfo {pages} {016002} (\bibinfo {year} {2022})}\BibitemShut
  {NoStop}%
\bibitem [{\citenamefont {Odette}\ and\ \citenamefont
  {Nanstad}(2009)}]{Odette2009}%
  \BibitemOpen
  \bibfield  {author} {\bibinfo {author} {\bibfnamefont {G.~R.}\ \bibnamefont
  {Odette}}\ and\ \bibinfo {author} {\bibfnamefont {R.~K.}\ \bibnamefont
  {Nanstad}},\ }\bibfield  {title} {\bibinfo {title} {Predictive reactor
  pressure vessel steel irradiation embrittlement models: Issues and
  opportunities},\ }\href {https://doi.org/10.1007/s11837-009-0097-4}
  {\bibfield  {journal} {\bibinfo  {journal} {JOM}\ }\textbf {\bibinfo {volume}
  {67}},\ \bibinfo {pages} {17} (\bibinfo {year} {2009})}\BibitemShut {NoStop}%
\bibitem [{\citenamefont {Knaster}\ \emph {et~al.}(2016)\citenamefont
  {Knaster}, \citenamefont {Moeslang},\ and\ \citenamefont
  {Muroga}}]{Knaster2016}%
  \BibitemOpen
  \bibfield  {author} {\bibinfo {author} {\bibfnamefont {J.}~\bibnamefont
  {Knaster}}, \bibinfo {author} {\bibfnamefont {A.}~\bibnamefont {Moeslang}},\
  and\ \bibinfo {author} {\bibfnamefont {T.}~\bibnamefont {Muroga}},\
  }\bibfield  {title} {\bibinfo {title} {Materials research for fusion},\
  }\href {https://doi.org/10.1038/nphys3735} {\bibfield  {journal} {\bibinfo
  {journal} {Nature Physics}\ }\textbf {\bibinfo {volume} {12}},\ \bibinfo
  {pages} {424} (\bibinfo {year} {2016})}\BibitemShut {NoStop}%
\bibitem [{\citenamefont {Was}(2015)}]{Was2015}%
  \BibitemOpen
  \bibfield  {author} {\bibinfo {author} {\bibfnamefont {G.}~\bibnamefont
  {Was}},\ }\bibfield  {title} {\bibinfo {title} {Challenges to the use of ion
  irradiation for emulating reactor irradiation},\ }\href
  {https://doi.org/10.1557/jmr.2015.73} {\bibfield  {journal} {\bibinfo
  {journal} {Journal of Materials Research}\ }\textbf {\bibinfo {volume}
  {30}},\ \bibinfo {pages} {1158} (\bibinfo {year} {2015})}\BibitemShut
  {NoStop}%
\bibitem [{\citenamefont {Sand}\ \emph {et~al.}(2013)\citenamefont {Sand},
  \citenamefont {Dudarev},\ and\ \citenamefont {Nordlund}}]{Sand2013}%
  \BibitemOpen
  \bibfield  {author} {\bibinfo {author} {\bibfnamefont {A.~E.}\ \bibnamefont
  {Sand}}, \bibinfo {author} {\bibfnamefont {S.~L.}\ \bibnamefont {Dudarev}},\
  and\ \bibinfo {author} {\bibfnamefont {K.}~\bibnamefont {Nordlund}},\
  }\bibfield  {title} {\bibinfo {title} {High-energy collision cascades in
  tungsten: Dislocation loops structure and clustering scaling laws},\ }\href
  {https://doi.org/10.1209/0295-5075/103/46003} {\bibfield  {journal} {\bibinfo
   {journal} {EPL}\ }\textbf {\bibinfo {volume} {103}},\ \bibinfo {pages}
  {46003} (\bibinfo {year} {2013})}\BibitemShut {NoStop}%
\bibitem [{\citenamefont {Nordlund}\ \emph
  {et~al.}(2018{\natexlab{a}})\citenamefont {Nordlund}, \citenamefont {Zinkle},
  \citenamefont {Sand}, \citenamefont {Granberg}, \citenamefont {Averback},
  \citenamefont {Stoller}, \citenamefont {Suzudo}, \citenamefont {Malerba},
  \citenamefont {Banhart}, \citenamefont {Weber} \emph
  {et~al.}}]{nordlund2018improving}%
  \BibitemOpen
  \bibfield  {author} {\bibinfo {author} {\bibfnamefont {K.}~\bibnamefont
  {Nordlund}}, \bibinfo {author} {\bibfnamefont {S.~J.}\ \bibnamefont
  {Zinkle}}, \bibinfo {author} {\bibfnamefont {A.~E.}\ \bibnamefont {Sand}},
  \bibinfo {author} {\bibfnamefont {F.}~\bibnamefont {Granberg}}, \bibinfo
  {author} {\bibfnamefont {R.~S.}\ \bibnamefont {Averback}}, \bibinfo {author}
  {\bibfnamefont {R.}~\bibnamefont {Stoller}}, \bibinfo {author} {\bibfnamefont
  {T.}~\bibnamefont {Suzudo}}, \bibinfo {author} {\bibfnamefont
  {L.}~\bibnamefont {Malerba}}, \bibinfo {author} {\bibfnamefont
  {F.}~\bibnamefont {Banhart}}, \bibinfo {author} {\bibfnamefont {W.~J.}\
  \bibnamefont {Weber}}, \emph {et~al.},\ }\bibfield  {title} {\bibinfo {title}
  {Improving atomic displacement and replacement calculations with physically
  realistic damage models},\ }\href
  {https://doi.org/https://doi.org/10.1038/s41467-018-03415-5} {\bibfield
  {journal} {\bibinfo  {journal} {Nature Communications}\ }\textbf {\bibinfo
  {volume} {9}},\ \bibinfo {pages} {1} (\bibinfo {year}
  {2018}{\natexlab{a}})}\BibitemShut {NoStop}%
\bibitem [{\citenamefont {Yang}\ and\ \citenamefont
  {Olsson}(2021)}]{yang2021full}%
  \BibitemOpen
  \bibfield  {author} {\bibinfo {author} {\bibfnamefont {Q.}~\bibnamefont
  {Yang}}\ and\ \bibinfo {author} {\bibfnamefont {P.}~\bibnamefont {Olsson}},\
  }\bibfield  {title} {\bibinfo {title} {Full energy range primary radiation
  damage model},\ }\href
  {https://doi.org/https://doi.org/10.1103/PhysRevMaterials.5.073602}
  {\bibfield  {journal} {\bibinfo  {journal} {Physical Review Materials}\
  }\textbf {\bibinfo {volume} {5}},\ \bibinfo {pages} {073602} (\bibinfo {year}
  {2021})}\BibitemShut {NoStop}%
\bibitem [{\citenamefont {Yi}\ \emph {et~al.}(2015)\citenamefont {Yi},
  \citenamefont {Sand}, \citenamefont {Mason}, \citenamefont {Kirk},
  \citenamefont {Roberts}, \citenamefont {Nordlund},\ and\ \citenamefont
  {Dudarev}}]{Yi2015}%
  \BibitemOpen
  \bibfield  {author} {\bibinfo {author} {\bibfnamefont {X.}~\bibnamefont
  {Yi}}, \bibinfo {author} {\bibfnamefont {A.~E.}\ \bibnamefont {Sand}},
  \bibinfo {author} {\bibfnamefont {D.~R.}\ \bibnamefont {Mason}}, \bibinfo
  {author} {\bibfnamefont {M.~A.}\ \bibnamefont {Kirk}}, \bibinfo {author}
  {\bibfnamefont {S.~G.}\ \bibnamefont {Roberts}}, \bibinfo {author}
  {\bibfnamefont {K.}~\bibnamefont {Nordlund}},\ and\ \bibinfo {author}
  {\bibfnamefont {S.~L.}\ \bibnamefont {Dudarev}},\ }\bibfield  {title}
  {\bibinfo {title} {Direct observation of size scaling and elastic interaction
  between nano-scale defects in collision cascades},\ }\href
  {https://doi.org/10.1209/0295-5075/110/36001} {\bibfield  {journal} {\bibinfo
   {journal} {EPL}\ }\textbf {\bibinfo {volume} {110}},\ \bibinfo {pages}
  {36001} (\bibinfo {year} {2015})}\BibitemShut {NoStop}%
\bibitem [{\citenamefont {Oddershede}\ \emph {et~al.}(1993)\citenamefont
  {Oddershede}, \citenamefont {Dimon},\ and\ \citenamefont
  {Bohr}}]{Oddershede1993}%
  \BibitemOpen
  \bibfield  {author} {\bibinfo {author} {\bibfnamefont {L.}~\bibnamefont
  {Oddershede}}, \bibinfo {author} {\bibfnamefont {P.}~\bibnamefont {Dimon}},\
  and\ \bibinfo {author} {\bibfnamefont {J.}~\bibnamefont {Bohr}},\ }\bibfield
  {title} {\bibinfo {title} {{Self-Organized Criticality in Fragmenting}},\
  }\href {https://doi.org/10.1103/PhysRevLett.71.3107} {\bibfield  {journal}
  {\bibinfo  {journal} {Physical Review Letters}\ }\textbf {\bibinfo {volume}
  {71}},\ \bibinfo {pages} {3107} (\bibinfo {year} {1993})}\BibitemShut
  {NoStop}%
\bibitem [{\citenamefont {Granberg}\ \emph {et~al.}(2016)\citenamefont
  {Granberg}, \citenamefont {Nordlund}, \citenamefont {Ullah}, \citenamefont
  {Jin}, \citenamefont {Lu}, \citenamefont {Bei}, \citenamefont {Wang},
  \citenamefont {Djurabekova}, \citenamefont {Weber},\ and\ \citenamefont
  {Zhang}}]{granberg2016mechanism}%
  \BibitemOpen
  \bibfield  {author} {\bibinfo {author} {\bibfnamefont {F.}~\bibnamefont
  {Granberg}}, \bibinfo {author} {\bibfnamefont {K.}~\bibnamefont {Nordlund}},
  \bibinfo {author} {\bibfnamefont {M.~W.}\ \bibnamefont {Ullah}}, \bibinfo
  {author} {\bibfnamefont {K.}~\bibnamefont {Jin}}, \bibinfo {author}
  {\bibfnamefont {C.}~\bibnamefont {Lu}}, \bibinfo {author} {\bibfnamefont
  {H.}~\bibnamefont {Bei}}, \bibinfo {author} {\bibfnamefont {L.~M.}\
  \bibnamefont {Wang}}, \bibinfo {author} {\bibfnamefont {F.}~\bibnamefont
  {Djurabekova}}, \bibinfo {author} {\bibfnamefont {W.~J.}\ \bibnamefont
  {Weber}},\ and\ \bibinfo {author} {\bibfnamefont {Y.}~\bibnamefont {Zhang}},\
  }\bibfield  {title} {\bibinfo {title} {{Mechanism of Radiation Damage
  Reduction in Equiatomic Multicomponent Single Phase Alloys}},\ }\href
  {https://doi.org/10.1103/PhysRevLett.116.135504} {\bibfield  {journal}
  {\bibinfo  {journal} {Physical Review Letters}\ }\textbf {\bibinfo {volume}
  {116}},\ \bibinfo {pages} {135504} (\bibinfo {year} {2016})}\BibitemShut
  {NoStop}%
\bibitem [{\citenamefont {Derlet}\ and\ \citenamefont
  {Dudarev}(2020)}]{Derlet2020}%
  \BibitemOpen
  \bibfield  {author} {\bibinfo {author} {\bibfnamefont {P.~M.}\ \bibnamefont
  {Derlet}}\ and\ \bibinfo {author} {\bibfnamefont {S.~L.}\ \bibnamefont
  {Dudarev}},\ }\bibfield  {title} {\bibinfo {title} {Microscopic structure of
  a heavily irradiated material},\ }\href
  {https://doi.org/10.1103/PhysRevMaterials.4.023605} {\bibfield  {journal}
  {\bibinfo  {journal} {Physical Review Materials}\ }\textbf {\bibinfo {volume}
  {4}},\ \bibinfo {pages} {023605} (\bibinfo {year} {2020})}\BibitemShut
  {NoStop}%
\bibitem [{\citenamefont {Mason}\ \emph {et~al.}(2020)\citenamefont {Mason},
  \citenamefont {Das}, \citenamefont {Derlet}, \citenamefont {Dudarev},
  \citenamefont {London}, \citenamefont {Yu}, \citenamefont {Phillips},
  \citenamefont {Yang}, \citenamefont {Mizohata}, \citenamefont {Xu},\ and\
  \citenamefont {Hofmann}}]{Mason2020}%
  \BibitemOpen
  \bibfield  {author} {\bibinfo {author} {\bibfnamefont {D.~R.}\ \bibnamefont
  {Mason}}, \bibinfo {author} {\bibfnamefont {S.}~\bibnamefont {Das}}, \bibinfo
  {author} {\bibfnamefont {P.~M.}\ \bibnamefont {Derlet}}, \bibinfo {author}
  {\bibfnamefont {S.~L.}\ \bibnamefont {Dudarev}}, \bibinfo {author}
  {\bibfnamefont {A.~J.}\ \bibnamefont {London}}, \bibinfo {author}
  {\bibfnamefont {H.}~\bibnamefont {Yu}}, \bibinfo {author} {\bibfnamefont
  {N.~W.}\ \bibnamefont {Phillips}}, \bibinfo {author} {\bibfnamefont
  {D.}~\bibnamefont {Yang}}, \bibinfo {author} {\bibfnamefont {K.}~\bibnamefont
  {Mizohata}}, \bibinfo {author} {\bibfnamefont {R.}~\bibnamefont {Xu}},\ and\
  \bibinfo {author} {\bibfnamefont {F.}~\bibnamefont {Hofmann}},\ }\bibfield
  {title} {\bibinfo {title} {Observation of transient and asymptotic driven
  structural states of tungsten exposed to radiation},\ }\href
  {https://doi.org/10.1103/PhysRevLett.125.225503} {\bibfield  {journal}
  {\bibinfo  {journal} {Physical Review Letters}\ }\textbf {\bibinfo {volume}
  {125}},\ \bibinfo {pages} {225503} (\bibinfo {year} {2020})}\BibitemShut
  {NoStop}%
\bibitem [{\citenamefont {Mason}\ \emph {et~al.}(2021)\citenamefont {Mason},
  \citenamefont {Granberg}, \citenamefont {Boleininger}, \citenamefont
  {Schwarz-Selinger}, \citenamefont {Nordlund},\ and\ \citenamefont
  {Dudarev}}]{mason2021parameter}%
  \BibitemOpen
  \bibfield  {author} {\bibinfo {author} {\bibfnamefont {D.~R.}\ \bibnamefont
  {Mason}}, \bibinfo {author} {\bibfnamefont {F.}~\bibnamefont {Granberg}},
  \bibinfo {author} {\bibfnamefont {M.}~\bibnamefont {Boleininger}}, \bibinfo
  {author} {\bibfnamefont {T.}~\bibnamefont {Schwarz-Selinger}}, \bibinfo
  {author} {\bibfnamefont {K.}~\bibnamefont {Nordlund}},\ and\ \bibinfo
  {author} {\bibfnamefont {S.~L.}\ \bibnamefont {Dudarev}},\ }\bibfield
  {title} {\bibinfo {title} {Parameter-free quantitative simulation of
  high-dose microstructure and hydrogen retention in ion-irradiated tungsten},\
  }\href {https://doi.org/10.1103/PhysRevMaterials.5.095403} {\bibfield
  {journal} {\bibinfo  {journal} {Physical Review Materials}\ }\textbf
  {\bibinfo {volume} {5}},\ \bibinfo {pages} {095403} (\bibinfo {year}
  {2021})}\BibitemShut {NoStop}%
\bibitem [{\citenamefont {Warwick}\ \emph {et~al.}(2021)\citenamefont
  {Warwick}, \citenamefont {Boleininger},\ and\ \citenamefont
  {Dudarev}}]{Warwick2021}%
  \BibitemOpen
  \bibfield  {author} {\bibinfo {author} {\bibfnamefont {A.~R.}\ \bibnamefont
  {Warwick}}, \bibinfo {author} {\bibfnamefont {M.}~\bibnamefont
  {Boleininger}},\ and\ \bibinfo {author} {\bibfnamefont {S.~L.}\ \bibnamefont
  {Dudarev}},\ }\bibfield  {title} {\bibinfo {title} {Microstructural
  complexity and dimensional changes in heavily irradiated zirconium},\ }\href
  {https://doi.org/10.1103/PhysRevMaterials.5.113604} {\bibfield  {journal}
  {\bibinfo  {journal} {Physical Review Materials}\ }\textbf {\bibinfo {volume}
  {5}},\ \bibinfo {pages} {113604} (\bibinfo {year} {2021})}\BibitemShut
  {NoStop}%
\bibitem [{\citenamefont {Boleininger}\ \emph {et~al.}(2022)\citenamefont
  {Boleininger}, \citenamefont {Dudarev}, \citenamefont {Mason},\ and\
  \citenamefont {Mart{\'\i}nez}}]{boleininger2022volume}%
  \BibitemOpen
  \bibfield  {author} {\bibinfo {author} {\bibfnamefont {M.}~\bibnamefont
  {Boleininger}}, \bibinfo {author} {\bibfnamefont {S.~L.}\ \bibnamefont
  {Dudarev}}, \bibinfo {author} {\bibfnamefont {D.~R.}\ \bibnamefont {Mason}},\
  and\ \bibinfo {author} {\bibfnamefont {E.}~\bibnamefont {Mart{\'\i}nez}},\
  }\bibfield  {title} {\bibinfo {title} {Volume of a dislocation network},\
  }\href {https://doi.org/10.1103/PhysRevMaterials.6.063601} {\bibfield
  {journal} {\bibinfo  {journal} {Physical Review Materials}\ }\textbf
  {\bibinfo {volume} {6}},\ \bibinfo {pages} {063601} (\bibinfo {year}
  {2022})}\BibitemShut {NoStop}%
\bibitem [{\citenamefont {Hirst}\ \emph {et~al.}(2022)\citenamefont {Hirst},
  \citenamefont {Granberg}, \citenamefont {Kombaiah}, \citenamefont {Cao},
  \citenamefont {Middlemas}, \citenamefont {Kemp}, \citenamefont {Li},
  \citenamefont {Nordlund},\ and\ \citenamefont {Short}}]{hirst2022revealing}%
  \BibitemOpen
  \bibfield  {author} {\bibinfo {author} {\bibfnamefont {C.~A.}\ \bibnamefont
  {Hirst}}, \bibinfo {author} {\bibfnamefont {F.}~\bibnamefont {Granberg}},
  \bibinfo {author} {\bibfnamefont {B.}~\bibnamefont {Kombaiah}}, \bibinfo
  {author} {\bibfnamefont {P.}~\bibnamefont {Cao}}, \bibinfo {author}
  {\bibfnamefont {S.}~\bibnamefont {Middlemas}}, \bibinfo {author}
  {\bibfnamefont {R.~S.}\ \bibnamefont {Kemp}}, \bibinfo {author}
  {\bibfnamefont {J.}~\bibnamefont {Li}}, \bibinfo {author} {\bibfnamefont
  {K.}~\bibnamefont {Nordlund}},\ and\ \bibinfo {author} {\bibfnamefont
  {M.~P.}\ \bibnamefont {Short}},\ }\bibfield  {title} {\bibinfo {title}
  {Revealing hidden defects through stored energy measurements of radiation
  damage},\ }\href {https://doi.org/https://doi.org/10.1126/sciadv.abn2733}
  {\bibfield  {journal} {\bibinfo  {journal} {Science advances}\ }\textbf
  {\bibinfo {volume} {8}},\ \bibinfo {pages} {eabn2733} (\bibinfo {year}
  {2022})}\BibitemShut {NoStop}%
\bibitem [{\citenamefont {Swinburne}\ \emph {et~al.}(2016)\citenamefont
  {Swinburne}, \citenamefont {Arakawa}, \citenamefont {Mori}, \citenamefont
  {Yasuda}, \citenamefont {Isshiki}, \citenamefont {Mimura}, \citenamefont
  {Uchikoshi},\ and\ \citenamefont {Dudarev}}]{swinburne2016fast}%
  \BibitemOpen
  \bibfield  {author} {\bibinfo {author} {\bibfnamefont {T.~D.}\ \bibnamefont
  {Swinburne}}, \bibinfo {author} {\bibfnamefont {K.}~\bibnamefont {Arakawa}},
  \bibinfo {author} {\bibfnamefont {H.}~\bibnamefont {Mori}}, \bibinfo {author}
  {\bibfnamefont {H.}~\bibnamefont {Yasuda}}, \bibinfo {author} {\bibfnamefont
  {M.}~\bibnamefont {Isshiki}}, \bibinfo {author} {\bibfnamefont
  {K.}~\bibnamefont {Mimura}}, \bibinfo {author} {\bibfnamefont
  {M.}~\bibnamefont {Uchikoshi}},\ and\ \bibinfo {author} {\bibfnamefont
  {S.~L.}\ \bibnamefont {Dudarev}},\ }\bibfield  {title} {\bibinfo {title}
  {Fast, vacancy-free climb of prismatic dislocation loops in bcc metals},\
  }\href {https://doi.org/https://doi.org/10.1038/srep30596} {\bibfield
  {journal} {\bibinfo  {journal} {Scientific reports}\ }\textbf {\bibinfo
  {volume} {6}},\ \bibinfo {pages} {30596} (\bibinfo {year}
  {2016})}\BibitemShut {NoStop}%
\bibitem [{\citenamefont {Rovelli}\ \emph {et~al.}(2017)\citenamefont
  {Rovelli}, \citenamefont {Dudarev},\ and\ \citenamefont
  {Sutton}}]{rovelli2017non}%
  \BibitemOpen
  \bibfield  {author} {\bibinfo {author} {\bibfnamefont {I.}~\bibnamefont
  {Rovelli}}, \bibinfo {author} {\bibfnamefont {S.~L.}\ \bibnamefont
  {Dudarev}},\ and\ \bibinfo {author} {\bibfnamefont {A.~P.}\ \bibnamefont
  {Sutton}},\ }\bibfield  {title} {\bibinfo {title} {Non-local model for
  diffusion-mediated dislocation climb and cavity growth},\ }\href
  {https://doi.org/https://doi.org/10.1016/j.jmps.2017.03.008} {\bibfield
  {journal} {\bibinfo  {journal} {Journal of the Mechanics and Physics of
  Solids}\ }\textbf {\bibinfo {volume} {103}},\ \bibinfo {pages} {121}
  (\bibinfo {year} {2017})}\BibitemShut {NoStop}%
\bibitem [{\citenamefont {Nordlund}\ \emph
  {et~al.}(2018{\natexlab{b}})\citenamefont {Nordlund}, \citenamefont {Zinkle},
  \citenamefont {Sand}, \citenamefont {Granberg}, \citenamefont {Averback},
  \citenamefont {Stoller}, \citenamefont {Suzudo}, \citenamefont {Malerba},
  \citenamefont {Banhart}, \citenamefont {Weber}, \citenamefont {Willaime},
  \citenamefont {Dudarev},\ and\ \citenamefont {Simeone}}]{Nordlund_JNM2018}%
  \BibitemOpen
  \bibfield  {author} {\bibinfo {author} {\bibfnamefont {K.}~\bibnamefont
  {Nordlund}}, \bibinfo {author} {\bibfnamefont {S.~J.}\ \bibnamefont
  {Zinkle}}, \bibinfo {author} {\bibfnamefont {A.~E.}\ \bibnamefont {Sand}},
  \bibinfo {author} {\bibfnamefont {F.}~\bibnamefont {Granberg}}, \bibinfo
  {author} {\bibfnamefont {R.~S.}\ \bibnamefont {Averback}}, \bibinfo {author}
  {\bibfnamefont {R.~E.}\ \bibnamefont {Stoller}}, \bibinfo {author}
  {\bibfnamefont {T.}~\bibnamefont {Suzudo}}, \bibinfo {author} {\bibfnamefont
  {L.}~\bibnamefont {Malerba}}, \bibinfo {author} {\bibfnamefont
  {F.}~\bibnamefont {Banhart}}, \bibinfo {author} {\bibfnamefont {W.~J.}\
  \bibnamefont {Weber}}, \bibinfo {author} {\bibfnamefont {F.}~\bibnamefont
  {Willaime}}, \bibinfo {author} {\bibfnamefont {S.~L.}\ \bibnamefont
  {Dudarev}},\ and\ \bibinfo {author} {\bibfnamefont {D.}~\bibnamefont
  {Simeone}},\ }\bibfield  {title} {\bibinfo {title} {Primary radiation damage:
  A review of current understanding and models},\ }\href
  {https://doi.org/https://doi.org/10.1016/j.jnucmat.2018.10.027} {\bibfield
  {journal} {\bibinfo  {journal} {Journal of Nuclear Materials}\ }\textbf
  {\bibinfo {volume} {512}},\ \bibinfo {pages} {450} (\bibinfo {year}
  {2018}{\natexlab{b}})}\BibitemShut {NoStop}%
\bibitem [{\citenamefont {Norgett}\ \emph {et~al.}(1975)\citenamefont
  {Norgett}, \citenamefont {Robinson},\ and\ \citenamefont
  {Torrens}}]{norgett1975proposed}%
  \BibitemOpen
  \bibfield  {author} {\bibinfo {author} {\bibfnamefont {M.~J.}\ \bibnamefont
  {Norgett}}, \bibinfo {author} {\bibfnamefont {M.~T.}\ \bibnamefont
  {Robinson}},\ and\ \bibinfo {author} {\bibfnamefont {I.~M.}\ \bibnamefont
  {Torrens}},\ }\bibfield  {title} {\bibinfo {title} {A proposed method of
  calculating displacement dose rates},\ }\href
  {https://doi.org/https://doi.org/10.1016/0029-5493(75)90035-7} {\bibfield
  {journal} {\bibinfo  {journal} {Nuclear Engineering and Design}\ }\textbf
  {\bibinfo {volume} {33}},\ \bibinfo {pages} {50} (\bibinfo {year}
  {1975})}\BibitemShut {NoStop}%
\bibitem [{\citenamefont {Nordlund}\ \emph {et~al.}(2006)\citenamefont
  {Nordlund}, \citenamefont {Wallenius},\ and\ \citenamefont
  {Malerba}}]{nordlund2006molecular}%
  \BibitemOpen
  \bibfield  {author} {\bibinfo {author} {\bibfnamefont {K.}~\bibnamefont
  {Nordlund}}, \bibinfo {author} {\bibfnamefont {J.}~\bibnamefont
  {Wallenius}},\ and\ \bibinfo {author} {\bibfnamefont {L.}~\bibnamefont
  {Malerba}},\ }\bibfield  {title} {\bibinfo {title} {Molecular dynamics
  simulations of threshold displacement energies in {Fe}},\ }\href
  {https://doi.org/https://doi.org/10.1016/j.nimb.2006.01.003} {\bibfield
  {journal} {\bibinfo  {journal} {Nuclear Instruments and Methods in Physics
  Research Section B: Beam Interactions with Materials and Atoms}\ }\textbf
  {\bibinfo {volume} {246}},\ \bibinfo {pages} {322} (\bibinfo {year}
  {2006})}\BibitemShut {NoStop}%
\bibitem [{\citenamefont {Lindhard}\ \emph {et~al.}(1963)\citenamefont
  {Lindhard}, \citenamefont {Scharff},\ and\ \citenamefont
  {Schi{\o}tt}}]{lindhard1963range}%
  \BibitemOpen
  \bibfield  {author} {\bibinfo {author} {\bibfnamefont {J.}~\bibnamefont
  {Lindhard}}, \bibinfo {author} {\bibfnamefont {M.}~\bibnamefont {Scharff}},\
  and\ \bibinfo {author} {\bibfnamefont {H.~E.}\ \bibnamefont {Schi{\o}tt}},\
  }\href@noop {} {\emph {\bibinfo {title} {Range concepts and heavy ion
  ranges}}}\ (\bibinfo  {publisher} {Munksgaard Copenhagen},\ \bibinfo {year}
  {1963})\BibitemShut {NoStop}%
\bibitem [{Note1()}]{Note1}%
  \BibitemOpen
  \bibinfo {note} {The energy required to melt a crystal with initial
  temperature $T_0$ is formally obtained by $E_\protect \mathrm {melt}^0 =
  c_\protect \mathrm {V} (T_\protect \mathrm {melt} - T_0) + L$\cite
  {de2016subcascade}, where $c_\protect \mathrm {V} \approx 3 k_\protect
  \mathrm {B}$ is molar heat capacity, $T_\protect \mathrm {melt}$ is melting
  temperature, and $L$ is latent heat of fusion per atom. In practice this
  value overestimates the melt size as it does not account for kinetic energy
  being rapidly conducted out of the cascade region into the surrounding
  crystal.}\BibitemShut {Stop}%
\bibitem [{\citenamefont {Stukowski}(2009)}]{stukowski2009visualization}%
  \BibitemOpen
  \bibfield  {author} {\bibinfo {author} {\bibfnamefont {A.}~\bibnamefont
  {Stukowski}},\ }\bibfield  {title} {\bibinfo {title} {Visualization and
  analysis of atomistic simulation data with {OVITO}--the open visualization
  tool},\ }\href
  {https://doi.org/https://doi.org/10.1088/0965-0393/18/1/015012} {\bibfield
  {journal} {\bibinfo  {journal} {Modelling and Simulation in Materials Science
  and Engineering}\ }\textbf {\bibinfo {volume} {18}},\ \bibinfo {pages}
  {015012} (\bibinfo {year} {2009})}\BibitemShut {NoStop}%
\bibitem [{\citenamefont {Dudarev}(2008)}]{Dudarev2008}%
  \BibitemOpen
  \bibfield  {author} {\bibinfo {author} {\bibfnamefont {S.~L.}\ \bibnamefont
  {Dudarev}},\ }\bibfield  {title} {\bibinfo {title} {{The non-Arrhenius
  migration of interstitial defects in bcc transition metals}},\ }\href
  {https://doi.org/10.1016/j.crhy.2007.09.019} {\bibfield  {journal} {\bibinfo
  {journal} {Comptes Rendus Physique}\ }\textbf {\bibinfo {volume} {9}},\
  \bibinfo {pages} {409 } (\bibinfo {year} {2008})}\BibitemShut {NoStop}%
\bibitem [{\citenamefont {Konobeyev}\ \emph {et~al.}(2017)\citenamefont
  {Konobeyev}, \citenamefont {Fischer}, \citenamefont {Korovin},\ and\
  \citenamefont {Simakov}}]{konobeyev2017evaluation}%
  \BibitemOpen
  \bibfield  {author} {\bibinfo {author} {\bibfnamefont {A.~Y.}\ \bibnamefont
  {Konobeyev}}, \bibinfo {author} {\bibfnamefont {U.}~\bibnamefont {Fischer}},
  \bibinfo {author} {\bibfnamefont {Y.~A.}\ \bibnamefont {Korovin}},\ and\
  \bibinfo {author} {\bibfnamefont {S.~P.}\ \bibnamefont {Simakov}},\
  }\bibfield  {title} {\bibinfo {title} {Evaluation of effective threshold
  displacement energies and other data required for the calculation of advanced
  atomic displacement cross-sections},\ }\href
  {https://doi.org/https://doi.org/10.1016/j.nucet.2017.08.007} {\bibfield
  {journal} {\bibinfo  {journal} {Nuclear Energy and Technology}\ }\textbf
  {\bibinfo {volume} {3}},\ \bibinfo {pages} {169} (\bibinfo {year}
  {2017})}\BibitemShut {NoStop}%
\bibitem [{\citenamefont {Mason}\ \emph {et~al.}(2019)\citenamefont {Mason},
  \citenamefont {Nguyen-Manh}, \citenamefont {Marinica}, \citenamefont
  {Alexander}, \citenamefont {Sand},\ and\ \citenamefont
  {Dudarev}}]{mason2019relaxation}%
  \BibitemOpen
  \bibfield  {author} {\bibinfo {author} {\bibfnamefont {D.~R.}\ \bibnamefont
  {Mason}}, \bibinfo {author} {\bibfnamefont {D.}~\bibnamefont {Nguyen-Manh}},
  \bibinfo {author} {\bibfnamefont {M.-C.}\ \bibnamefont {Marinica}}, \bibinfo
  {author} {\bibfnamefont {R.}~\bibnamefont {Alexander}}, \bibinfo {author}
  {\bibfnamefont {A.~E.}\ \bibnamefont {Sand}},\ and\ \bibinfo {author}
  {\bibfnamefont {S.~L.}\ \bibnamefont {Dudarev}},\ }\bibfield  {title}
  {\bibinfo {title} {Relaxation volumes of microscopic and mesoscopic
  irradiation-induced defects in tungsten},\ }\href
  {https://doi.org/10.1063/1.5094852} {\bibfield  {journal} {\bibinfo
  {journal} {Journal of Applied Physics}\ }\textbf {\bibinfo {volume} {126}},\
  \bibinfo {pages} {075112} (\bibinfo {year} {2019})}\BibitemShut {NoStop}%
\bibitem [{\citenamefont {Tian}\ \emph {et~al.}(2021)\citenamefont {Tian},
  \citenamefont {Wang}, \citenamefont {Feng}, \citenamefont {Zheng},
  \citenamefont {Liu},\ and\ \citenamefont {Zhou}}]{tian2021heavy}%
  \BibitemOpen
  \bibfield  {author} {\bibinfo {author} {\bibfnamefont {J.}~\bibnamefont
  {Tian}}, \bibinfo {author} {\bibfnamefont {H.}~\bibnamefont {Wang}}, \bibinfo
  {author} {\bibfnamefont {Q.}~\bibnamefont {Feng}}, \bibinfo {author}
  {\bibfnamefont {J.}~\bibnamefont {Zheng}}, \bibinfo {author} {\bibfnamefont
  {X.}~\bibnamefont {Liu}},\ and\ \bibinfo {author} {\bibfnamefont
  {W.}~\bibnamefont {Zhou}},\ }\bibfield  {title} {\bibinfo {title} {Heavy
  radiation damage in alpha zirconium at cryogenic temperature: {A}
  computational study},\ }\href
  {https://doi.org/https://doi.org/10.1016/j.jnucmat.2021.153159} {\bibfield
  {journal} {\bibinfo  {journal} {Journal of Nuclear Materials}\ }\textbf
  {\bibinfo {volume} {555}},\ \bibinfo {pages} {153159} (\bibinfo {year}
  {2021})}\BibitemShut {NoStop}%
\bibitem [{\citenamefont {Zhang}\ \emph {et~al.}(2021)\citenamefont {Zhang},
  \citenamefont {Zhang}, \citenamefont {Qiao},\ and\ \citenamefont
  {Wang}}]{zhang2021comparing}%
  \BibitemOpen
  \bibfield  {author} {\bibinfo {author} {\bibfnamefont {H.}~\bibnamefont
  {Zhang}}, \bibinfo {author} {\bibfnamefont {X.}~\bibnamefont {Zhang}},
  \bibinfo {author} {\bibfnamefont {L.}~\bibnamefont {Qiao}},\ and\ \bibinfo
  {author} {\bibfnamefont {P.}~\bibnamefont {Wang}},\ }\bibfield  {title}
  {\bibinfo {title} {Comparing deuterium retention in heavy ion damaged
  tungsten measured by glow discharge optical emission spectroscopy, nuclear
  reaction analysis and thermal desorption spectroscopy},\ }\href
  {https://doi.org/https://doi.org/10.1088/1402-4896/ac1381} {\bibfield
  {journal} {\bibinfo  {journal} {Physica Scripta}\ }\textbf {\bibinfo {volume}
  {96}},\ \bibinfo {pages} {115602} (\bibinfo {year} {2021})}\BibitemShut
  {NoStop}%
\bibitem [{\citenamefont {Mason}\ \emph {et~al.}(2017)\citenamefont {Mason},
  \citenamefont {Nguyen-Manh},\ and\ \citenamefont
  {Becquart}}]{mason2017empirical}%
  \BibitemOpen
  \bibfield  {author} {\bibinfo {author} {\bibfnamefont {D.~R.}\ \bibnamefont
  {Mason}}, \bibinfo {author} {\bibfnamefont {D.}~\bibnamefont {Nguyen-Manh}},\
  and\ \bibinfo {author} {\bibfnamefont {C.~S.}\ \bibnamefont {Becquart}},\
  }\bibfield  {title} {\bibinfo {title} {An empirical potential for simulating
  vacancy clusters in tungsten},\ }\href
  {https://doi.org/https://doi.org/10.1088/1361-648X/aa9776} {\bibfield
  {journal} {\bibinfo  {journal} {Journal of Physics: Condensed Matter}\
  }\textbf {\bibinfo {volume} {29}},\ \bibinfo {pages} {505501} (\bibinfo
  {year} {2017})}\BibitemShut {NoStop}%
\bibitem [{\citenamefont {Ackland}\ and\ \citenamefont
  {Vitek}(1990)}]{ackland1990many}%
  \BibitemOpen
  \bibfield  {author} {\bibinfo {author} {\bibfnamefont {G.~J.}\ \bibnamefont
  {Ackland}}\ and\ \bibinfo {author} {\bibfnamefont {V.}~\bibnamefont
  {Vitek}},\ }\bibfield  {title} {\bibinfo {title} {Many-body potentials and
  atomic-scale relaxations in noble-metal alloys},\ }\href
  {https://doi.org/https://doi.org/10.1103/PhysRevB.41.10324} {\bibfield
  {journal} {\bibinfo  {journal} {Physical Review B}\ }\textbf {\bibinfo
  {volume} {41}},\ \bibinfo {pages} {10324} (\bibinfo {year}
  {1990})}\BibitemShut {NoStop}%
\bibitem [{\citenamefont {Mendelev}\ and\ \citenamefont
  {Ackland}(2007)}]{Mendelev2007}%
  \BibitemOpen
  \bibfield  {author} {\bibinfo {author} {\bibfnamefont {M.~I.}\ \bibnamefont
  {Mendelev}}\ and\ \bibinfo {author} {\bibfnamefont {G.~J.}\ \bibnamefont
  {Ackland}},\ }\bibfield  {title} {\bibinfo {title} {{Development of an
  interatomic potential for the simulation of phase transformations in
  zirconium}},\ }\href {https://doi.org/10.1080/09500830701191393} {\bibfield
  {journal} {\bibinfo  {journal} {Philosophical Magazine Letters}\ }\textbf
  {\bibinfo {volume} {87}},\ \bibinfo {pages} {349} (\bibinfo {year}
  {2007})}\BibitemShut {NoStop}%
\bibitem [{\citenamefont {Wielunska}\ \emph {et~al.}(2020)\citenamefont
  {Wielunska}, \citenamefont {Mayer}, \citenamefont {Schwarz-Selinger},
  \citenamefont {Sand},\ and\ \citenamefont {Jacob}}]{wielunska2020deuterium}%
  \BibitemOpen
  \bibfield  {author} {\bibinfo {author} {\bibfnamefont {B.}~\bibnamefont
  {Wielunska}}, \bibinfo {author} {\bibfnamefont {M.}~\bibnamefont {Mayer}},
  \bibinfo {author} {\bibfnamefont {T.}~\bibnamefont {Schwarz-Selinger}},
  \bibinfo {author} {\bibfnamefont {A.~E.}\ \bibnamefont {Sand}},\ and\
  \bibinfo {author} {\bibfnamefont {W.}~\bibnamefont {Jacob}},\ }\bibfield
  {title} {\bibinfo {title} {Deuterium retention in tungsten irradiated by
  different ions},\ }\href
  {https://doi.org/https://doi.org/10.1088/1741-4326/ab9a65} {\bibfield
  {journal} {\bibinfo  {journal} {Nuclear Fusion}\ }\textbf {\bibinfo {volume}
  {60}},\ \bibinfo {pages} {096002} (\bibinfo {year} {2020})}\BibitemShut
  {NoStop}%
\bibitem [{\citenamefont {Tyburska}\ \emph {et~al.}(2009)\citenamefont
  {Tyburska}, \citenamefont {Alimov}, \citenamefont {Ogorodnikova},
  \citenamefont {Schmid},\ and\ \citenamefont {Ertl}}]{tyburska2009deuterium}%
  \BibitemOpen
  \bibfield  {author} {\bibinfo {author} {\bibfnamefont {B.}~\bibnamefont
  {Tyburska}}, \bibinfo {author} {\bibfnamefont {V.~K.}\ \bibnamefont
  {Alimov}}, \bibinfo {author} {\bibfnamefont {O.~V.}\ \bibnamefont
  {Ogorodnikova}}, \bibinfo {author} {\bibfnamefont {K.}~\bibnamefont
  {Schmid}},\ and\ \bibinfo {author} {\bibfnamefont {K.}~\bibnamefont {Ertl}},\
  }\bibfield  {title} {\bibinfo {title} {Deuterium retention in self-damaged
  tungsten},\ }\href
  {https://doi.org/https://doi.org/10.1016/j.jnucmat.2009.10.046} {\bibfield
  {journal} {\bibinfo  {journal} {Journal of Nuclear Materials}\ }\textbf
  {\bibinfo {volume} {395}},\ \bibinfo {pages} {150} (\bibinfo {year}
  {2009})}\BibitemShut {NoStop}%
\bibitem [{\citenamefont {'t~Hoen}\ \emph {et~al.}(2012)\citenamefont
  {'t~Hoen}, \citenamefont {Tyburska-P\"{u}schel}, \citenamefont {Ertl},
  \citenamefont {Mayer}, \citenamefont {Rapp}, \citenamefont {Kleijn},\ and\
  \citenamefont {Zeijlmans~van Emmichoven}}]{t2012saturation}%
  \BibitemOpen
  \bibfield  {author} {\bibinfo {author} {\bibfnamefont {M.~H.~J.}\
  \bibnamefont {'t~Hoen}}, \bibinfo {author} {\bibfnamefont {B.}~\bibnamefont
  {Tyburska-P\"{u}schel}}, \bibinfo {author} {\bibfnamefont {K.}~\bibnamefont
  {Ertl}}, \bibinfo {author} {\bibfnamefont {M.}~\bibnamefont {Mayer}},
  \bibinfo {author} {\bibfnamefont {J.}~\bibnamefont {Rapp}}, \bibinfo {author}
  {\bibfnamefont {A.~W.}\ \bibnamefont {Kleijn}},\ and\ \bibinfo {author}
  {\bibfnamefont {P.~A.}\ \bibnamefont {Zeijlmans~van Emmichoven}},\ }\bibfield
   {title} {\bibinfo {title} {Saturation of deuterium retention in self-damaged
  tungsten exposed to high-flux plasmas},\ }\bibfield  {journal} {\bibinfo
  {journal} {Nuclear Fusion}\ }\textbf {\bibinfo {volume} {52}},\ \href
  {https://doi.org/https://doi.org/10.1088/0029-5515/52/2/023008}
  {https://doi.org/10.1088/0029-5515/52/2/023008} (\bibinfo {year}
  {2012})\BibitemShut {NoStop}%
\bibitem [{\citenamefont {M{\"o}ller}\ \emph {et~al.}(2020)\citenamefont
  {M{\"o}ller}, \citenamefont {Krug}, \citenamefont {Rayaprolu}, \citenamefont
  {Kuhn}, \citenamefont {Jou{\ss}en},\ and\ \citenamefont
  {Kreter}}]{moller2020deuterium}%
  \BibitemOpen
  \bibfield  {author} {\bibinfo {author} {\bibfnamefont {S.}~\bibnamefont
  {M{\"o}ller}}, \bibinfo {author} {\bibfnamefont {R.}~\bibnamefont {Krug}},
  \bibinfo {author} {\bibfnamefont {R.}~\bibnamefont {Rayaprolu}}, \bibinfo
  {author} {\bibfnamefont {B.}~\bibnamefont {Kuhn}}, \bibinfo {author}
  {\bibfnamefont {E.}~\bibnamefont {Jou{\ss}en}},\ and\ \bibinfo {author}
  {\bibfnamefont {A.}~\bibnamefont {Kreter}},\ }\bibfield  {title} {\bibinfo
  {title} {Deuterium retention in tungsten and reduced activation steels after
  3 {MeV} proton irradiation},\ }\href
  {https://doi.org/https://doi.org/10.1016/j.nme.2020.100742} {\bibfield
  {journal} {\bibinfo  {journal} {Nuclear Materials and Energy}\ }\textbf
  {\bibinfo {volume} {23}},\ \bibinfo {pages} {100742} (\bibinfo {year}
  {2020})}\BibitemShut {NoStop}%
\bibitem [{\citenamefont {Meslin}\ \emph {et~al.}(2010)\citenamefont {Meslin},
  \citenamefont {Lambrecht}, \citenamefont {Hernández-Mayoral}, \citenamefont
  {Bergner}, \citenamefont {Malerba}, \citenamefont {Pareige}, \citenamefont
  {Radiguet}, \citenamefont {Barbu}, \citenamefont {Gómez-Briceño},
  \citenamefont {Ulbricht},\ and\ \citenamefont {Almazouzi}}]{Meslin2010}%
  \BibitemOpen
  \bibfield  {author} {\bibinfo {author} {\bibfnamefont {E.}~\bibnamefont
  {Meslin}}, \bibinfo {author} {\bibfnamefont {M.}~\bibnamefont {Lambrecht}},
  \bibinfo {author} {\bibfnamefont {M.}~\bibnamefont {Hernández-Mayoral}},
  \bibinfo {author} {\bibfnamefont {F.}~\bibnamefont {Bergner}}, \bibinfo
  {author} {\bibfnamefont {L.}~\bibnamefont {Malerba}}, \bibinfo {author}
  {\bibfnamefont {P.}~\bibnamefont {Pareige}}, \bibinfo {author} {\bibfnamefont
  {B.}~\bibnamefont {Radiguet}}, \bibinfo {author} {\bibfnamefont
  {A.}~\bibnamefont {Barbu}}, \bibinfo {author} {\bibfnamefont
  {D.}~\bibnamefont {Gómez-Briceño}}, \bibinfo {author} {\bibfnamefont
  {A.}~\bibnamefont {Ulbricht}},\ and\ \bibinfo {author} {\bibfnamefont
  {A.}~\bibnamefont {Almazouzi}},\ }\bibfield  {title} {\bibinfo {title}
  {{Characterization of neutron-irradiated ferritic model alloys and a RPV
  steel from combined APT, SANS, TEM and PAS analyses}},\ }\href
  {https://doi.org/10.1016/j.jnucmat.2009.12.021} {\bibfield  {journal}
  {\bibinfo  {journal} {Journal of Nuclear Materials}\ }\textbf {\bibinfo
  {volume} {406}},\ \bibinfo {pages} {73} (\bibinfo {year} {2010})}\BibitemShut
  {NoStop}%
\bibitem [{\citenamefont {Heinola}\ \emph {et~al.}(2010)\citenamefont
  {Heinola}, \citenamefont {Ahlgren}, \citenamefont {Nordlund},\ and\
  \citenamefont {Keinonen}}]{heinola2010hydrogen}%
  \BibitemOpen
  \bibfield  {author} {\bibinfo {author} {\bibfnamefont {K.}~\bibnamefont
  {Heinola}}, \bibinfo {author} {\bibfnamefont {T.}~\bibnamefont {Ahlgren}},
  \bibinfo {author} {\bibfnamefont {K.}~\bibnamefont {Nordlund}},\ and\
  \bibinfo {author} {\bibfnamefont {J.}~\bibnamefont {Keinonen}},\ }\bibfield
  {title} {\bibinfo {title} {Hydrogen interaction with point defects in
  tungsten},\ }\href
  {https://doi.org/https://doi.org/10.1103/PhysRevB.82.094102} {\bibfield
  {journal} {\bibinfo  {journal} {Physical Review B}\ }\textbf {\bibinfo
  {volume} {82}},\ \bibinfo {pages} {094102} (\bibinfo {year}
  {2010})}\BibitemShut {NoStop}%
\bibitem [{\citenamefont {Keys}\ \emph {et~al.}(1968)\citenamefont {Keys},
  \citenamefont {Smith},\ and\ \citenamefont {Moteff}}]{keys1968high}%
  \BibitemOpen
  \bibfield  {author} {\bibinfo {author} {\bibfnamefont {L.~K.}\ \bibnamefont
  {Keys}}, \bibinfo {author} {\bibfnamefont {J.~P.}\ \bibnamefont {Smith}},\
  and\ \bibinfo {author} {\bibfnamefont {J.}~\bibnamefont {Moteff}},\
  }\bibfield  {title} {\bibinfo {title} {High-temperature recovery of tungsten
  after neutron irradiation},\ }\href
  {https://doi.org/https://doi.org/10.1103/PhysRev.176.851} {\bibfield
  {journal} {\bibinfo  {journal} {Physical Review}\ }\textbf {\bibinfo {volume}
  {176}},\ \bibinfo {pages} {851} (\bibinfo {year} {1968})}\BibitemShut
  {NoStop}%
\bibitem [{\citenamefont {Shimada}\ \emph {et~al.}(2014)\citenamefont
  {Shimada}, \citenamefont {Cao}, \citenamefont {Otsuka}, \citenamefont {Hara},
  \citenamefont {Kobayashi}, \citenamefont {Oya},\ and\ \citenamefont
  {Hatano}}]{shimada2014irradiation}%
  \BibitemOpen
  \bibfield  {author} {\bibinfo {author} {\bibfnamefont {M.}~\bibnamefont
  {Shimada}}, \bibinfo {author} {\bibfnamefont {G.}~\bibnamefont {Cao}},
  \bibinfo {author} {\bibfnamefont {T.}~\bibnamefont {Otsuka}}, \bibinfo
  {author} {\bibfnamefont {M.}~\bibnamefont {Hara}}, \bibinfo {author}
  {\bibfnamefont {M.}~\bibnamefont {Kobayashi}}, \bibinfo {author}
  {\bibfnamefont {Y.}~\bibnamefont {Oya}},\ and\ \bibinfo {author}
  {\bibfnamefont {Y.}~\bibnamefont {Hatano}},\ }\bibfield  {title} {\bibinfo
  {title} {Irradiation effect on deuterium behaviour in low-dose {HFIR}
  neutron-irradiated tungsten},\ }\href
  {https://doi.org/https://doi.org/10.1088/0029-5515/55/1/013008} {\bibfield
  {journal} {\bibinfo  {journal} {Nuclear Fusion}\ }\textbf {\bibinfo {volume}
  {55}},\ \bibinfo {pages} {013008} (\bibinfo {year} {2014})}\BibitemShut
  {NoStop}%
\bibitem [{\citenamefont {Huang}\ \emph {et~al.}(2018)\citenamefont {Huang},
  \citenamefont {Gilbert},\ and\ \citenamefont {Marian}}]{huang2018simulating}%
  \BibitemOpen
  \bibfield  {author} {\bibinfo {author} {\bibfnamefont {C.-H.}\ \bibnamefont
  {Huang}}, \bibinfo {author} {\bibfnamefont {M.~R.}\ \bibnamefont {Gilbert}},\
  and\ \bibinfo {author} {\bibfnamefont {J.}~\bibnamefont {Marian}},\
  }\bibfield  {title} {\bibinfo {title} {Simulating irradiation hardening in
  tungsten under fast neutron irradiation including {Re} production by
  transmutation},\ }\href
  {https://doi.org/https://doi.org/10.1016/j.jnucmat.2017.11.026} {\bibfield
  {journal} {\bibinfo  {journal} {Journal of Nuclear Materials}\ }\textbf
  {\bibinfo {volume} {499}},\ \bibinfo {pages} {204} (\bibinfo {year}
  {2018})}\BibitemShut {NoStop}%
\bibitem [{\citenamefont {Plimpton}(1995)}]{plimpton1995fast}%
  \BibitemOpen
  \bibfield  {author} {\bibinfo {author} {\bibfnamefont {S.}~\bibnamefont
  {Plimpton}},\ }\bibfield  {title} {\bibinfo {title} {Fast parallel algorithms
  for short-range molecular dynamics},\ }\href
  {https://doi.org/https://doi.org/10.1006/jcph.1995.1039} {\bibfield
  {journal} {\bibinfo  {journal} {Journal of computational physics}\ }\textbf
  {\bibinfo {volume} {117}},\ \bibinfo {pages} {1} (\bibinfo {year}
  {1995})}\BibitemShut {NoStop}%
\bibitem [{\citenamefont {Ziegler}(2004)}]{ziegler2004srim}%
  \BibitemOpen
  \bibfield  {author} {\bibinfo {author} {\bibfnamefont {J.~F.}\ \bibnamefont
  {Ziegler}},\ }\bibfield  {title} {\bibinfo {title} {{SRIM-2003}},\ }\href
  {https://doi.org/https://doi.org/10.1016/j.nimb.2004.01.208} {\bibfield
  {journal} {\bibinfo  {journal} {Nuclear Instruments and Methods in Physics
  Research Section B: Beam Interactions with Materials and Atoms}\ }\textbf
  {\bibinfo {volume} {219}},\ \bibinfo {pages} {1027} (\bibinfo {year}
  {2004})}\BibitemShut {NoStop}%
\bibitem [{\citenamefont {Mason}(2015)}]{mason2015incorporating}%
  \BibitemOpen
  \bibfield  {author} {\bibinfo {author} {\bibfnamefont {D.~R.}\ \bibnamefont
  {Mason}},\ }\bibfield  {title} {\bibinfo {title} {Incorporating non-adiabatic
  effects in embedded atom potentials for radiation damage cascade
  simulations},\ }\href
  {https://doi.org/https://doi.org/10.1088/0953-8984/27/14/145401} {\bibfield
  {journal} {\bibinfo  {journal} {Journal of Physics: Condensed Matter}\
  }\textbf {\bibinfo {volume} {27}},\ \bibinfo {pages} {145401} (\bibinfo
  {year} {2015})}\BibitemShut {NoStop}%
\bibitem [{\citenamefont {De~Backer}\ \emph {et~al.}(2016)\citenamefont
  {De~Backer}, \citenamefont {Sand}, \citenamefont {Nordlund}, \citenamefont
  {Luneville}, \citenamefont {Simeone},\ and\ \citenamefont
  {Dudarev}}]{de2016subcascade}%
  \BibitemOpen
  \bibfield  {author} {\bibinfo {author} {\bibfnamefont {A.}~\bibnamefont
  {De~Backer}}, \bibinfo {author} {\bibfnamefont {A.~E.}\ \bibnamefont {Sand}},
  \bibinfo {author} {\bibfnamefont {K.}~\bibnamefont {Nordlund}}, \bibinfo
  {author} {\bibfnamefont {L.}~\bibnamefont {Luneville}}, \bibinfo {author}
  {\bibfnamefont {D.}~\bibnamefont {Simeone}},\ and\ \bibinfo {author}
  {\bibfnamefont {S.~L.}\ \bibnamefont {Dudarev}},\ }\bibfield  {title}
  {\bibinfo {title} {Subcascade formation and defect cluster size scaling in
  high-energy collision events in metals},\ }\href
  {https://doi.org/https://doi.org/10.1209/0295-5075/115/26001} {\bibfield
  {journal} {\bibinfo  {journal} {EPL (Europhysics Letters)}\ }\textbf
  {\bibinfo {volume} {115}},\ \bibinfo {pages} {26001} (\bibinfo {year}
  {2016})}\BibitemShut {NoStop}%
\bibitem [{csd()}]{csd3}%
  \BibitemOpen
  \href {www.csd3.cam.ac.uk} {\bibinfo {title}
  {www.csd3.cam.ac.uk}}\BibitemShut {NoStop}%
\end{thebibliography}%

\

\noindent\textbf{\sffamily  ACKNOWLEDGMENTS}\\
{\footnotesize
We are grateful to J.C. Haley, A.J. London, Q. Yang, F. Granberg, and P.M. Derlet for discussions that stimulated this study. This work has been carried out within the framework of the EUROfusion Consortium, funded by the European Union via the Euratom Research and Training Programme (Grant Agreement No. 101052200 - EUROfusion), and by the RCUK Energy Programme, Grant No. EP/W006839/1. To obtain further information on the data and models underlying the paper please contact PublicationsManager@ukaea.uk. The views and opinions expressed herein do not necessarily reflect those of the European Commission. The authors acknowledge the use of the Cambridge Service for Data Driven Discovery (CSD3) and associated support services provided by the University of Cambridge Research Computing Services \cite{csd3} that assisted the completion of this study.
}

\

\noindent\textbf{\sffamily AUTHOR CONTRIBUTIONS}\\
{\footnotesize
M.B. developed the concept and performed the simulations and their analysis. D.R.M. developed the method for determining vacancy content in bcc, fcc, and hcp crystals, and contributed to the development of the model for saturated vacancy content. A.E.S. contributed to data analysis and its interpretation in context of cascade simulations. S.L.D. contributed to developing the theoretical concept of the steady state. All authors contributed to the manuscript preparation and approved the final version of the manuscript.
}

\end{document}